\documentclass[reprint,amsmath,amssymb,aps,nofootinbib]{revtex4-2}
\usepackage{graphicx}
\usepackage{subfig}
\usepackage{dcolumn}
\usepackage{bm}
\usepackage{xcolor}

\begin{document}

\title{f-mode oscillations of anisotropic neutron stars in full general relativity}

\author{Sushovan Mondal}
\email{smondal@imsc.res.in}
\affiliation{The Institute of Mathematical Sciences, HBNI, Taramani, Chennai 600113, India.}
\affiliation{Homi Bhabha National Institute, Training School Complex, Anushakti Nagar, Mumbai 400094, India.}

\author{Manjari Bagchi}
 \email{manjari@imsc.res.in}
\affiliation{The Institute of Mathematical Sciences, HBNI, Taramani, Chennai 600113, India.}
\affiliation{Homi Bhabha National Institute, Training School Complex, Anushakti Nagar, Mumbai 400094, India.}

\date{\today}

\begin{abstract}
We investigate f-mode oscillations of anisotropic, non-rotating, neutron stars within the framework of full general relativity, considering linear order perturbations in both the metric and the fluid. First, we present the equations governing unperturbed stellar structures as well as oscillations under a phenomenological ansatz to account for local anisotropy. Then, we solve those equations for two different equations of states, namely  BSk21 and  BSk19, the former being stiffer than the latter. For both of the cases, we consider only stable neutron stars. We see that, moderately anisotropic neutron stars with the tangential pressure larger than the radial pressure, can give more massive stable neutron stars than the isotropic or highly anisotropic ones. We find that the frequency of the f-mode exhibits a linear relationship with the square root of the average density of neutron stars and the slope of the linear fit depends on the anisotropic strength. We also see that, for any given value of the anisotropic strength, neutron stars of higher masses have higher values of the frequency. For lower values of the mass, the increase of the frequency with the mass is linear, but for higher values of the mass, the frequency increases rapidly with the increase in the value of the mass. However, this non-linear rise in the frequency with the mass is not prominent when the radial pressure is larger than the tangential pressure. We also see that for a fixed value of a small mass, higher anisotropy leads to a larger value of the frequency, but when the fixed mass is above a threshold value, higher anisotropy leads to a smaller value of the frequency. Moreover, for a fixed mass of the neutron star and for the same amount of the anisotropy, the value of the frequency is higher for the softer equation of state, but the nature of the variation in the frequency with the change in the anisotropic strength is similar for the two equations of state. We also find that the damping time of the f-mode oscillation decreases as the mass of the neutron star increases for all values of the anisotropic strength. However, for mildly anisotropic neutron stars, a slight increase in the damping time occurs near the stable maximum mass. Moreover, for a fixed mass of the neutron star and for the same amount of the anisotropy, the value of the damping time is lower for the softer equation of state, but the nature of the variation in the damping time with the change in the anisotropic strength is similar for the two equations of state.

\end{abstract}

\maketitle

\section{\label{intro} Introduction}

Among the various celestial bodies in the universe, neutron stars stand out as some of the most exotic objects due to their extremely high densities. Consequently, general relativity plays a crucial role in describing the overall structure of a neutron star. Neutron stars, along with the less dense white dwarfs, belong to the compact star family. In modeling the structure and associated properties of neutron stars, it is typically assumed that the pressure within the star is isotropic. However, the possibility of anisotropic pressure cannot be ruled out. The concept of an anisotropic fluid, one with different pressures in the tangential and radial directions, was first introduced by \citet{Lemaitre:1933gd} in his study of the expansion of the universe. Afterward,  \citet{Bowers:1974tgi} studied the equilibrium configuration of relativistic compact stars with pressure anisotropy. They demonstrated that the local anisotropy in compact stars significantly affects observable properties such as mass and surface redshift.

The pressure anisotropy inside a neutron star can be triggered by various phenomena, e.g., superfluidity \cite{Ruderman:1972aj, Hoffberg1970, Sokolov80}, pion condensation \cite{Sawyer:1972cq}, skyrmionic interactions \cite{nelmes:2012}, etc. The presence of the viscosity can also induce local anisotropy inside compact stars \cite{barreto1992equation,barreto1993exploding} like neutron stars. \citet{Yazadjiev:2011ks} modeled magnetars in general relativity in a nonperturbative way, where the constituent fluid was considered anisotropic due to the presence of the magnetic field. Elasticity in compact stars like neutron stars can also be described by the local anisotropy \cite{alho2022compact, Karlovini:2002fc}. Additional studies related to anisotropic pressure inside self gravitating objects like neutron stars have been described in \citet{Herrera:1997plx} and the references therein.

Neutron stars are laboratories to study various theories of physics at high densities. These objects are observed across a wide range of the electromagnetic spectrum. However, there are several properties that cannot be probed through electromagnetic waves. The recent detections of gravitational waves have opened up a completely new window for the observation of neutron stars and other compact objects. So far, only the gravitational waves emitted during the late inspiral, the merger, and the ringdown phase of the components of binary systems of compact objects have been detected by ground based detectors \cite{LIGOScientific:2016aoc, LIGOScientific:2017vwq, LIGOScientific:2021qlt}. Recently, the Pulsar Timing Array collaboration has found the evidence of gravitational waves in the nanohertz regime \cite{agazie2023nanograv,antoniadis2023second,lee2023searching}. In the future, it is expected that detectors like LISA (Laser Interferometer Space Antenna), Einstein telescope, etc., would detect gravitational waves generated from various sources. The oscillations of neutron stars are one of the promising sources of gravitational waves. In the last phase of the life cycle of a massive star, a supernova explosion occurs, and the newly formed neutron star might oscillate violently. Additionally, if the merger of two compact stars forms a neutron star, that object would also oscillate. Part of the energy from the oscillation would be emitted as gravitational radiation, propagating as ripples of space-time or gravitational waves. Due to the emission of gravitational waves, the oscillation would be damped.  \citet{thorne_campolattaro_1967} first theoretically studied the non-radial oscillations of neutron stars within the framework of full general relativity. Later, \citet{Lindblom:1983ps} and \citet{Detweiler:1985zz} conducted similar calculations with some modifications. Using the same formalism, the non-radial oscillation modes are calculated for superfluid neutron stars \cite{Comer:1999rs}, quark stars \cite{Sotani:2003zc}, neutron stars with density discontinuity \cite{Sotani:2001bb, Miniutti:2002bh}, etc. Recent studies on the impact of the equation of state on the neutron star oscillation modes reveal the fact that the inter-nuclei interactions have important effects on the quasi-normal modes \cite{Kunjipurayil:2022zah}. Most of these studies assume pressure isotropy inside neutron stars. 

The effects of anisotropy on non-radial oscillation modes were first studied by \citet{hillebrandt1976anisotropic} in the Newtonian framework. This study revealed that pressure anisotropy plays a very important role in mode frequencies. Later, \citet{Doneva:2012rd} calculated the frequencies for the f-mode and the p-mode oscillations using the Cowling approximation, where they ignored the metric perturbation. In a recent study, \citet{Curi:2022nnt} extended this work for the case of realistic equations of state. All of the aforementioned studies showed that anisotropy in neutron stars can change the numerical value of the frequencies of the quasi-normal modes. Although the metric perturbation is ignored, the Cowling approximation provides sufficiently accurate results for fluid perturbations. However, the possibility of errors due to ignoring the metric perturbation might not be negligible. Recently, \citet{Sotani:2020mwc} found that the mode frequencies under the Cowling approximation are estimated with approximately $20\%$ accuracy. Another drawback of the Cowling approximation is that, as the metric perturbation is ignored, the energy loss due to gravitational waves is unaccounted for. In the full framework of general relativity that takes into account perturbations of the metric, the oscillation of the metric propagates as gravitational waves, which carry away the energy of the oscillations. As a result, the oscillation experiences damping, leading to a decrease in its amplitude over time. The characteristic time associated with this damping is known as the damping time and is mathematically defined at the end of Sec. \ref{subsec:perschemeNum}. The damping time cannot be calculated using the Cowling approximation.

A few recent studies have hinted that it might be possible to determine the frequencies of the f-mode oscillations of a neutron star observationally \cite{Afle:2023mab, Roy:2023gzi}. As these studies assumed isotropic neutron stars under the Cowling approximation, they did not address the damping time. However, similar observations could also lead to measurements of the damping times. As the extent of anisotropy affects both the frequency and the damping time of the f-mode oscillations, it might also be possible to quantify the anisotropy via such observations.

The motivation of the present work is to study the properties associated with the f-mode oscillations of anisotropic neutron stars within the framework of full general relativity. Our focus is mainly to derive the governing differential equations for f-mode oscillations of neutron stars with pressure anisotropy and to find the mode frequencies and the damping times. Moreover, we study only polar perturbations, characterized by the metric perturbation functions that exhibit an even parity under the spatial inversion ($(\theta,\phi) \rightarrow (\pi - \theta, \pi + \phi)$), where $r$, $\theta$, and $\phi$ denote coordinates in a spherical polar coordinate system. Out of the various polar modes, we focus on the fundamental (f) modes. We have considered mainly BSk21 equation of state, which connects the radial pressure with the density inside the neutron stars. To describe the tangential pressure, we have considered a phenomenological ansatz described by \citet{Horvat:2010xf}. We show how the f-mode frequency and the associated damping time change due to the presence of the pressure anisotropy, for various masses of neutron stars, as well as various extents of the anisotropy inside those.

This paper is organized as follows. In Sec. \ref{sec:equilib}, we detail the equilibrium model of anisotropic neutron stars. This involves presenting the modified Tolman-Oppenheimer-Volkoff (TOV) equations due to anisotropy, describing the equation of state (EoS), and outlining the anisotropy ansatz. Moving to Sec. \ref{sec:perscheme}, we delve into the perturbation scheme for neutron stars. This section covers both of the analytical and numerical techniques employed to compute the frequencies and the damping times of the f-mode oscillations. In Sec. \ref{sec:results}, we present our results. Finally, we conclude the paper in Sec. \ref{sec:conclusion} with a summary and a discussion of our findings. In this paper, all mathematical expressions are written in the natural units, setting $G=1$, $c=1$, where $G$ is the gravitational constant and $c$ is the speed of light in vacuum. 

\section{Equilibrium configuration of anisotropic neutron stars in general relativity}\label{sec:equilib}
In the case of a spherically symmetric non-rotating space-time, the metric can be written in the well-known form
\begin{equation}
	ds^2 = -e^{\nu(r)}dt^2 + e^{\lambda(r)}dr^2 + r^2(d\theta^2+ \sin^2 \theta d\phi^2) , \label{metric}
\end{equation}
where $\nu(r)$ and $\lambda(r)$ are metric functions that depend on the radial coordinate ($r$) only. We take an anisotropic fluid as the matter source in the field equations, where the radial component of the pressure $(p_r)$ and the tangential component of the pressure $(p_t)$ are non-identical. Note that, like the metric functions, the radial and the tangential components of the pressure inside a neutron star also depend on $r$. For simplicity, we will not explicitly indicate the $r$ dependence from now on.

The anisotropy parameter is defined as,
\begin{equation}
	\chi = p_t - p_r .
	\label{eq:aniso_param}
\end{equation}
Note that, in this article, the anisotropy parameter $\chi$ is defined in the same way as done by \citet{Horvat:2010xf}, whereas some other studies \cite{Doneva:2012rd,Curi:2022nnt} define the anisotropy parameter as $(p_r - p_t)$. The energy momentum tensor of the anisotropic fluid can be written in the form \cite{Mak:2001eb,Arbanil:2016wud}
\begin{equation}
	T_{~\alpha}^{\beta} = (\rho  + p_t)u^\beta u_\alpha + \delta_{~\alpha}^{\beta} \, p_t + (p_r - p_t) s^\beta s_\alpha , \label{emt}
\end{equation}
where $\rho$ is the fluid matter density. The space-like radial unit vector $s^\alpha$ is defined as:
\begin{equation}
s^\alpha = (0,e^{-\lambda/2},0,0) ,
\end{equation}
and the fluid 4-velocity vector is given by:
\begin{equation}
u^\beta = (e^{-\nu /2},0,0,0) ,
\end{equation}
which satisfies the conditions:
\begin{subequations}
	\begin{align}
		u^{\alpha}u_{\alpha} &= -1 ,\\
		s^\alpha s_\alpha &= 1 ,\\
		s^\beta u_\beta &= 0 .
	\end{align}	
\end{subequations}	
The non-zero components of the energy-momentum tensor ($T_{~\alpha}^{\beta}$) are only $diag(-\rho, p_r, p_t, p_t)$. The space-time geometry and the matter distribution are related by Einstein equations:
\begin{equation}
	G_{~\alpha}^{\beta} = 8 \pi T_{~\alpha}^{\beta} , \label{einstein}
\end{equation}  
where $G_{~\alpha}^{\beta}$ is the Einstein tensor, describing the spacetime geometry. Using equations (\ref{metric}) and (\ref{emt}), the Einstein equations can be written as:
\begin{eqnarray}
	e^{-\lambda}\left(\frac{\lambda'}{r} - \frac{1}{r^2}\right) + \frac{1}{r^2} = 8 \pi \rho ,\label{eq1} \\
	e^{-\lambda}\left(\frac{\nu'}{r} + \frac{1}{r^2}\right) - \frac{1}{r^2} =  8 \pi p_r , \label{eq2} \\
	\frac{1}{2} e^{-\lambda}\left(\nu'' - \frac{1}{2} \nu'\lambda' + \frac{1}{2}{\nu'}^2 + \frac{\nu' - \lambda'}{r}\right) = 8 \pi p_t ,\label{eq3} 
\end{eqnarray}
where a prime denotes the differentiation with respect to $r$. From equations (\ref{eq1}), (\ref{eq2}), and (\ref{eq3}), we get the equation of hydrostatic equilibrium in the presence of the pressure anisotropy, which can be written as (in natural units): 
\begin{equation}
	p_r' = - \frac{\nu'}{2}(\rho  + p_r) + \frac{2\chi}{r}. \label{hydro}
\end{equation} 
The interior metric function ($r<R$, where $R$ is the radius of the star) can be found from Eq. (\ref{eq1}) as: 
\begin{equation}
	e^{-\lambda} = 1 - \frac{2 m}{r},\text{ where } m =4 \pi\int_0^r \rho(r') r'^2 dr'~, \label{inner}
\end{equation} is the mass enclosed within a spherical region of radius $r$ inside the star.
Using Eqs. (\ref{eq2}) and (\ref{inner}), the hydrostatic equilibrium equation (\ref{hydro}) can be written as: 
\begin{equation}
	p_r' = - \frac{(\rho + p_r) (m +4 \pi p_r r^3)}{r(r - 2 m)} + \frac{2\chi}{r}.
	\label{eq:modTOV}
\end{equation}
This is the modified Tolman-Oppenheimer-Volkoff (TOV) equation, which  accounts for the local pressure anisotropy. To solve this equation, we need to specify the equation of state of the neutron star matter, i.e., the dependence of $p_r$ on $\rho$ and the anisotropy parameter $\chi$. As the boundary conditions, we set a finite radial pressure at the center of the star and zero radial pressure at the surface of the star. This also implies zero density at the surface of the star. 

Additionally, the metric functions must match the exterior Schwarzschild metric at $r = R$, i.e.,
\begin{equation}
	\lambda(R) = -\nu(R) = -\ln \left(1 - \frac{2M}{R}\right),
\end{equation} 
where $M$ is the total mass of the star, i.e., $M = m(R)$.

\subsection{Description of the equation of state and the anisotropy parameter}\label{sec:eosAnisotropy}
As mentioned earlier, to construct a model of a neutron star, i.e., to solve the modified TOV equation (Eq.(\ref{eq:modTOV})), one needs to specify how the pressure inside the star varies with the density, which is known as the equation of state (EoS). There are studies of microscopic physics leading to various Equations of State (EsoS)\cite[and references therein]{Lattimer:2021emm, Burgio:2021vgk}.

However, those studies are usually for isotropic matter, and no rigorous study has been performed to model the pressure anisotropy inside a neutron star. Hence, people use different ansatzes for the anisotropy parameter $\chi$. One popular ansatz for $\chi$ was proposed by \citet{Horvat:2010xf} as: 
\begin{equation}
	\chi = \tau p_r \mu, \label{quasiparam}
\end{equation}
where $\mu = 2 m /r$ is the local compactness and $\tau$ is a parameter governing the strength of the anisotropy. The chosen form of the anisotropy parameter in Eq. (\ref{quasiparam}) possesses two appealing characteristics. First, the anisotropy parameter vanishes at the center of the star, as the compactness scales as $\mu \sim r^2$ when $r \to 0$, ensuring the regularity of the anisotropy parameter. Second, in the non-relativistic regime, where $\mu$ is significantly smaller than unity, $\chi$ also becomes very small, and hence like the original TOV equation, the modified TOV equation also agrees with the Newtonian pressure balance equation for hydrodstatic equilibrium. Additionally, being consistent with previous studies \cite{Doneva:2012rd, Folomeev:2018ioy, Silva:2014fca}, we consider the values of $\tau$ within the range $-2 \leq \tau \leq 2$. This range covers different theoretically motivated values that differ from each other within this range. As examples, \citet{Sawyer:1972cq} found that if the anisotropy is due to the pion condensation, then the range should be $0 \leq \tau \leq 1$ while \citet{nelmes:2012} found $\tau = -2 $ in their skyrmonic calculations.  

In the present work, we use the above ansatz for the anisotropy in association with the analytical representation of the Brussels - Montreal unified EoS for the nuclear matter, known as BSk19, BSk20, BSk21 \cite{Goriely:2010bm, Pearson:2011zz, Pearson:2012hz}, which model $p_r (\rho)$ with two parameters $\zeta$ and $\sigma$ that are parametrized as \cite{Potekhin:2013qqa}:

\begin{widetext}
\begin{equation}\label{BSk21}
	\begin{split}
		\zeta = \frac{a_1 + a_2 \sigma + a_3 \sigma^3}{1 + a_4 \sigma} f(a_5 (\sigma - a_6)) &+ (a_7 + a_8 \sigma) f(a_9 (a_6 - \sigma)) +  (a_{10} + a_{11} \sigma) f(a_{12}(a_{13} - \sigma)) \\
		&+(a_{14} + a_{15} \sigma)f(a_{16}(a_{17} - \sigma)) + \frac{a_{18}}{1 + [a_{19} (\sigma - a_{20})]^2} + \frac{a_{21}}{1 + [a_{22} (\sigma - a_{23})]^2},
	\end{split}
\end{equation}
\end{widetext}

where $\zeta = \log_{10}(p_r \, {\rm ~in ~ dyn~ cm^{-2}})$, $\sigma = \log_{10}(\rho \, {\rm ~ in ~ g ~ cm^{-3}})$, and $f(x) = [\exp(x) + 1]^{-1}$. The values of the coefficients $a_i (i = 1, \ldots , 23)$ are available in \citet{Potekhin:2013qqa}. Throughout this article, our analysis primarily uses BSk21 EoS. However, to compare our results across different levels of matter stiffness, we also consider BSk19 EoS in selected cases. 

Although one can solve the modified TOV equation for various values of the central density ($\rho_c$) and the anisotropic strength, but all of the resulting neutron stars would not be stable. First, to avoid the spontaneous collapse of the matter, the pressure should be monotonically nondecreasing function of the density \cite{RhoadesRuffini74}, i.e., $\partial p_r / \partial \rho >0$ and $\partial p_t / \partial \rho >0$. Second, the speed of sound can not be negative, i.e., $v_{sr} > 0$ and $v_{st} > 0$, where $v_{sr}$ and $v_{st}$ are the speed of sound in the radial and the tangential directions, respectively, defined as $v_{sr}^2 = \partial p_r / \partial \rho$ and $v_{st}^2 = \partial p_t / \partial \rho$. Any realistic EoS satisfy the condition $v_{sr} > 0$. However, as we do not have very rigorous studies on the pressure anisotropy, we rather use some plausible ansatzes, for a chosen set of the EoS and the anisotropy ansatz, the condition $v_{st} > 0$ might be violated for some (or all) values of the density and hence should be checked. Third, the causality condition should be satisfied, i.e., $v_{sr} < v_{\rm light}$ and $v_{st} < v_{\rm light}$, where $v_{\rm light}$ is the speed of light in the neutron star matter. As we do not know the value  of $v_{\rm light}$, we set it as $v_{\rm light} = 0.95 \, c$ where $c$ is the speed of light in vacuum (1 in the natural unit). Fourth, the neutron star must be stable under radial oscillation, which is possible only when $\partial M / \partial \rho_c >0$ \cite{Glendenning:1997wn}.   

\begin{figure}
	\centering
	\includegraphics[width=0.5\textwidth]{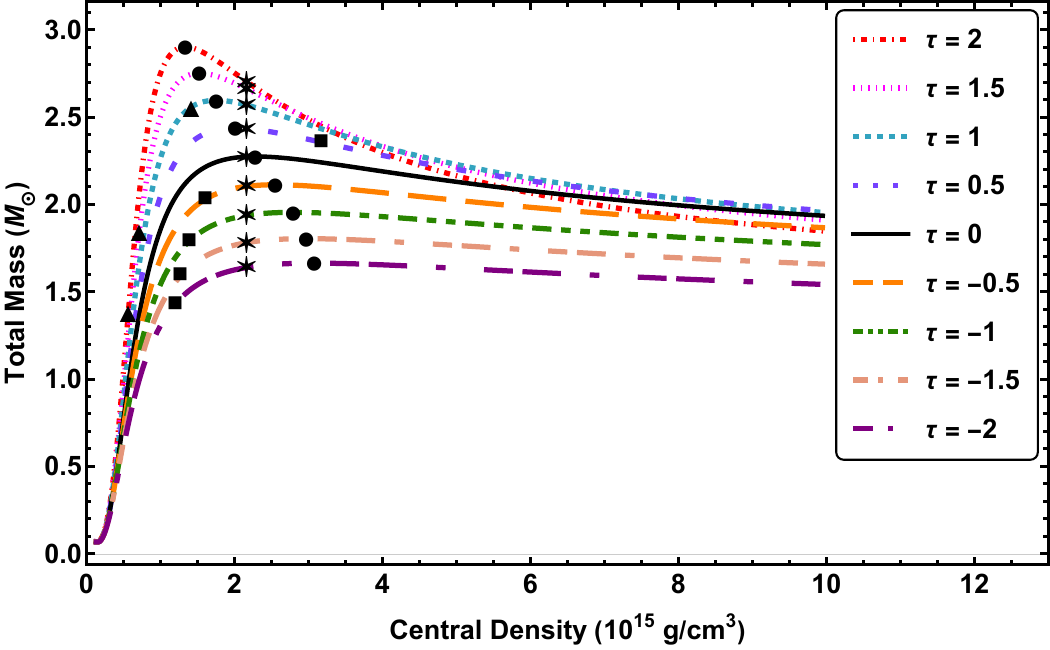}
	\caption{Mass ($M$)-Central density ($\rho_c$) profiles for anisotropic neutron stars with $-2 \leq\tau\leq 2$ for BSk21 EoS. Filled circles on each profile represent the points where $\partial M/\partial \rho_c = 0$, the asterisks represent the values of $\rho_c$ up to which $v_{sr} \leq 0.95$ throughout the star, the filled squares represent the values of $\rho_c$ up to which $v_{st} \leq 0.95$ throughout the star, and the filled triangles denote the values of $\rho_c$ up to which $v_{st} \geq 0$ throughout the star.}
	\label{mass-central-density}
\end{figure}

\begin{figure}
	\centering
	\includegraphics[width=0.5\textwidth]{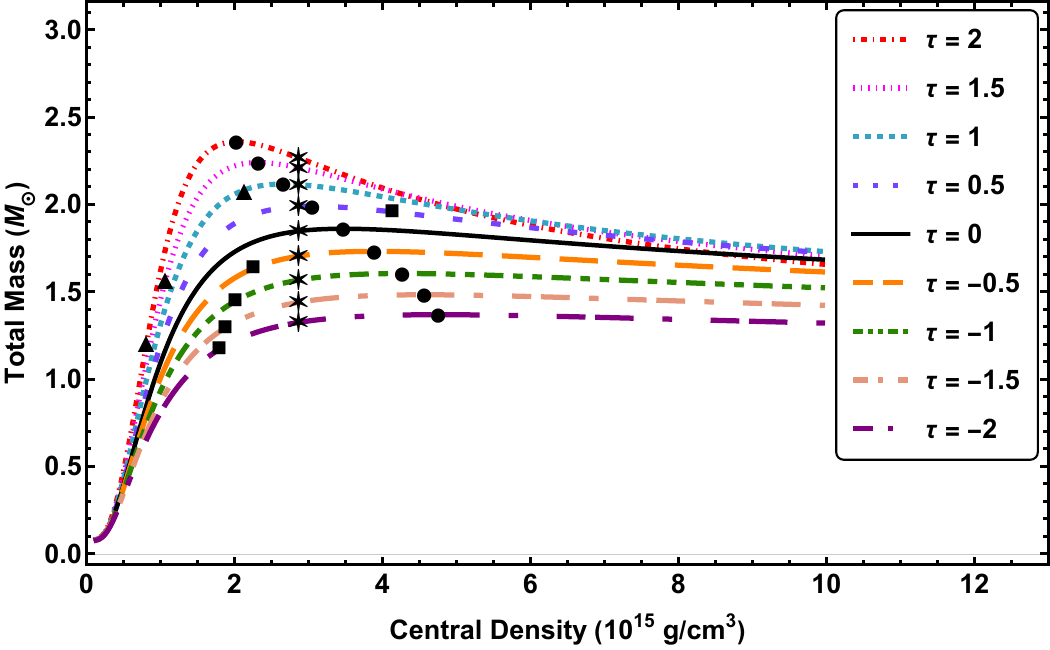}
	\caption{Mass ($M$)-Central density ($\rho_c$) profiles for anisotropic neutron stars with $-2 \leq\tau\leq 2$ for BSk19 EoS. Filled circles on each profile represent the points where $\partial M/\partial \rho_c = 0$, the asterisks stars represent the values of $\rho_c$ up to which $v_{sr} \leq 0.95$ throughout the star, the filled squares represent the values of $\rho_c$ up to which $v_{st} \leq 0.95$ throughout the star, and the filled triangles denote the values of $\rho_c$ up to which $v_{st} \geq 0$ throughout the star.}
	\label{mass-central-density-bsk19}
\end{figure}

To check the validity of our chosen EsoS, the anisotropy ansatz, and the range of the anisotropic strength $\tau$, in Figs. \ref{mass-central-density} and \ref{mass-central-density-bsk19}, we plot the values of the total mass ($M$) obtained by solving the modified TOV equation against the corresponding values of the central densities ($\rho_c$) for various values of $\tau$ (in the range of $-2$ to 2 as discussed earlier) for BSk21 and BSk19 EsoS respectively. In both of these figures, filled circles on each profile represent the points where $\partial M/\partial \rho_c = 0$, the asterisks mark the values of $\rho_c$ up to which $v_{sr} \leq 0.95$ throughout the star, the filled squares represent the values of $\rho_c$ up to which $v_{st} \leq 0.95$ throughout the star, and the filled triangles denote the values of $\rho_c$ up to which $v_{st} \geq 0$ throughout the star.

We see that, for both BSk21 and BSk19 EsoS, as the value of $\tau$ increases, the maximum value of the central density up to which we get neutron stars that are stable against radial oscillations decreases. Moreover, for $\tau \leq 0$, $v_{sr}$ reaches the value $0.95$ inside neutron stars at a value of $\rho_c$ that is smaller than the value that gives $\partial M / \partial \rho_c = 0$, and $v_{st}$ reach the limit of  $0.95$ at an even smaller value of $\rho_c$. The situation is the opposite when $\tau \geq 0$. For sufficiently large values of $\tau$, e.g., 1.5 and higher, $v_{st}$ never reaches the causal limit $0.95$. However, for such values of $\tau$, the value of $v_{st}$ does not remain positive throughout the star, except for very small values of $\rho_c$. On the other hand, for the chosen set of EsoS, the anisotropy ansatz, and the range of the values of $\tau$, $v_{sr}$ always remain positive throughout the star.

\begin{figure}
	\centering
	\includegraphics[width=0.5\textwidth]{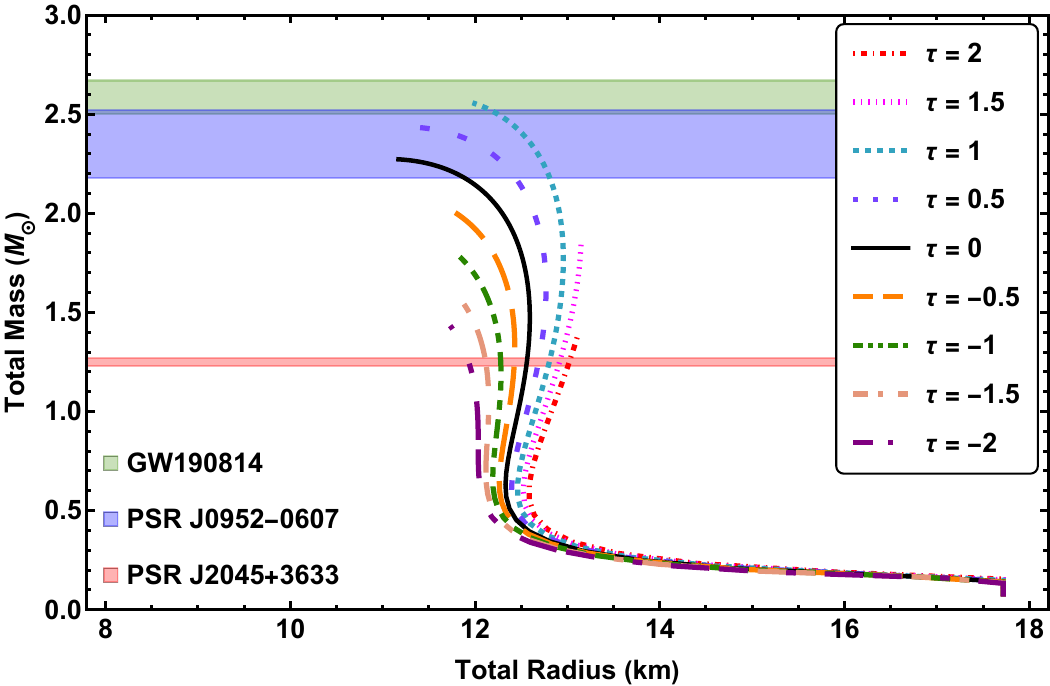}
	\caption{Mass-Radius profiles for anisotropic stable neutron stars with $-2 \leq\tau\leq 2$ and BSk21 EoS.}
	\label{MR}
\end{figure}

\begin{figure}
	\centering
	\includegraphics[width=0.5\textwidth]{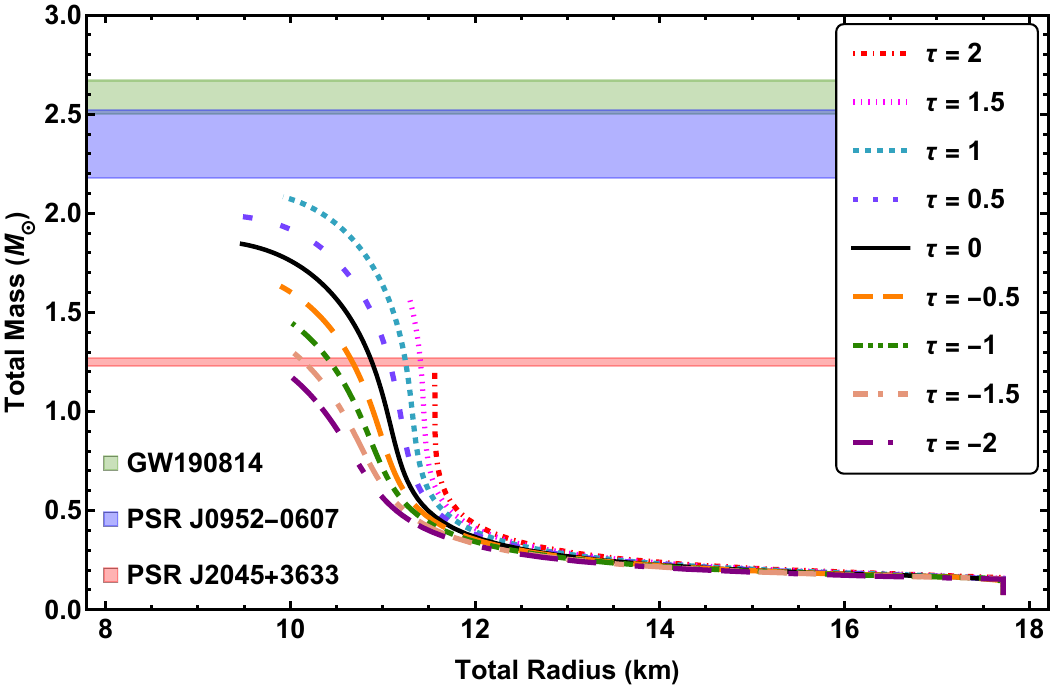}
	\caption{Mass-Radius profiles for anisotropic neutron stars with $-2 \leq\tau\leq 2$ and BSk19 EoS.}
	\label{MRbsk19}
\end{figure}

Next, we plot the mass$-$radius curves for various values of $\tau$ (again in the range of $-2$ to 2) for BSk21 EoS in Fig. \ref{MR} and for BSk19 EoS in Fig. \ref{MRbsk19}, both only up to the masses that satisfy all of the four stability conditions discussed earlier. In both of the figures, in addition to mass-radius curves, we have shown the measured mass ranges of PSR J2045$+$3633 \cite{McKee:2020pzp} and PSR J0952$-$0607 \cite{Romani:2022jhd}, representing a low-mass ($ \rm 1.252 \pm 0.021~M_\odot$) and a high mass ($\rm 2.35 \pm 0.17 ~M_\odot$) neutron star, respectively, as well as the band for the event GW190814 \cite{LIGOScientific:2020zkf}, corresponding to an object after merger in the mass range of $\rm 2.50 - 2.67~M_\odot$. This object lies in the so-called `mass gap', implying that it could be either a highly massive neutron star or a light black hole.

We see that for a range of the anisotropic strength, BSk21 EoS can lead to neutron stars massive enough to satisfy all three measurements. However, BSk19 EoS fails to produce masses higher than $2.1 ~ {\rm M_\odot}$. Hence, we prefer BSk21 EoS, although we can not rule out a softer EoS like BSk19 EoS. Especially, if we keep in mind the fact that it might be possible to obtain neutron stars heavier than $2.1 ~ {\rm M_\odot}$ with a different anisotropy ansatz, even for BSk19 EoS.

Note that, all our calculations, i.e., obtaining the stellar structure by solving the modified TOV equation (as presented in this section) and the study of the f-mode oscillations (will be presented in the next two sections), are done within the static limit, i.e., for non-rotating stars. Hence, we do not show the excluded region known as the `mass-shedding limit' for low mass neutron stars where the rotating neutron stars break due to centrifugal force. However, observed neutron stars are known to be fast rotators. In particular, the spin frequencies of PSR J2045$+$3633 and PSR J0952$-$0607 are 31.56 Hz \cite{McKee:2020pzp} and 707.31 Hz \cite{Bassa:2017zpe}, respectively. Low mass neutron stars ($\lesssim 0.5 ~{\rm M_{\odot}}$) with such high spin frequencies cannot exist in the nature.

The use of the static configuration does not affect the stellar structure significantly for the neutron stars of high masses. As an example, Bagchi \cite{Bagchi:2008su} showed that the maximum mass for a stable, non-rotating isotropic neutron star with APR EoS is 2.19 ${\rm M_{\odot}}$ whereas the maximum mass for a stable isotropic neutron stars rotating with a spin frequency of 796 Hz (with APR EoS) is 2.20 ${\rm M_{\odot}}$. This justifies our choice of studying even anisotropic neutron stars in the static limit. It will be interesting to study both the stellar structure and the f-mode oscillations of anisotropic neutron stars under full general relativistic framework including the rotation, in the future.

\begin{figure}
	\centering
	\includegraphics[width=0.5\textwidth]{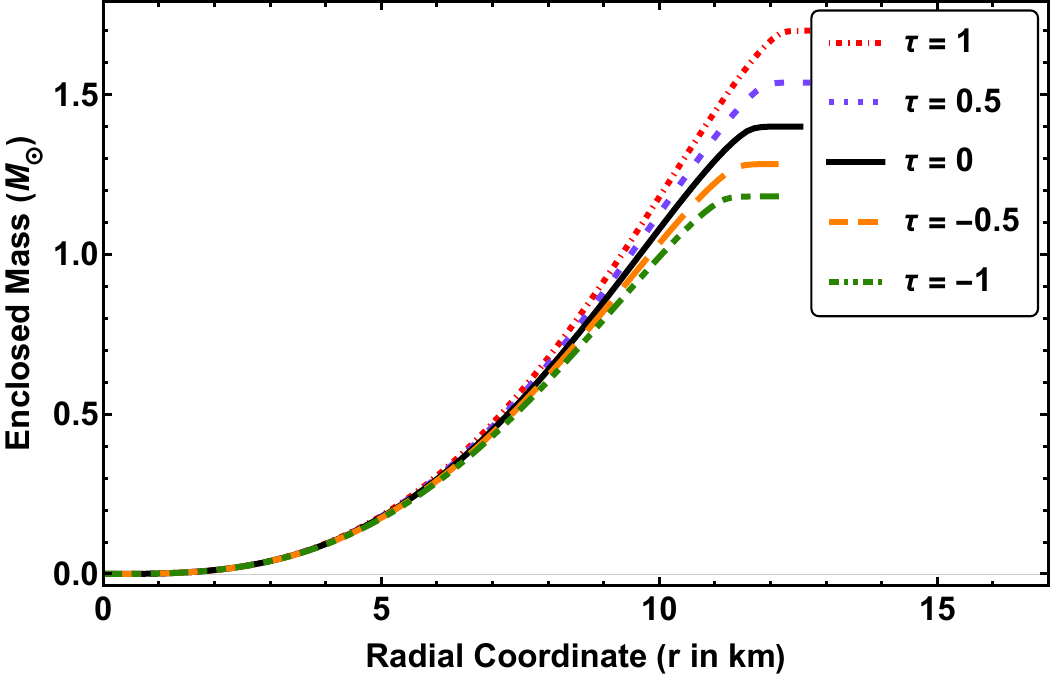}
	\caption{The variation of the enclosed mass with the radial coordinate inside neutron stars for various values of the anisotropic strength, each for $\rho_c = ~ 7.2955 \times 10^{14}~{\rm g \, cm^{-3}}$ and BSk21 EoS. }
	\label{massvar}
\end{figure}

In Figure \ref{massvar}, we plot the radial coordinate (in km) inside the neutron star along the abscissa, and the corresponding enclosed mass in the unit of solar mass along the ordinate for BSk21 EoS. Various curves in this figure represent different values of $\tau$, while the value of $\rho_c$ is taken as $7.2955 \times 10^{14}~{\rm g \, cm^{-3}}$ for all of the cases. This is the central density that gives a stable neutron star of a mass of $\rm M = 1.4~ {\rm M_\odot}$ for $\tau = 0$. However, this particular value of $\rho_c$ gives stable neutron stars only in the range of $ -1 \leq \tau \leq 1$. In particular, for $\tau = -1.0, -0.5, 0.5, 1.0$, we get neutron stars of mass $1.18 ~ {\rm M_\odot}$, $ 1.28 ~{\rm M_\odot}$, $ 1.53 ~{\rm M_\odot}$, and $ 1.69~{\rm M_\odot}$, respectively, for $\rho_c = 7.2955 \times 10^{14}~{\rm g \, cm^{-3}}$. Hence, whenever we choose $\rho_c = 7.2955 \times 10^{14}~{\rm g \, cm^{-3}}$, we restrict ourselves in the range of $ -1 \leq \tau \leq 1$.

\begin{figure}
	\centering
	\includegraphics[width=0.5\textwidth]{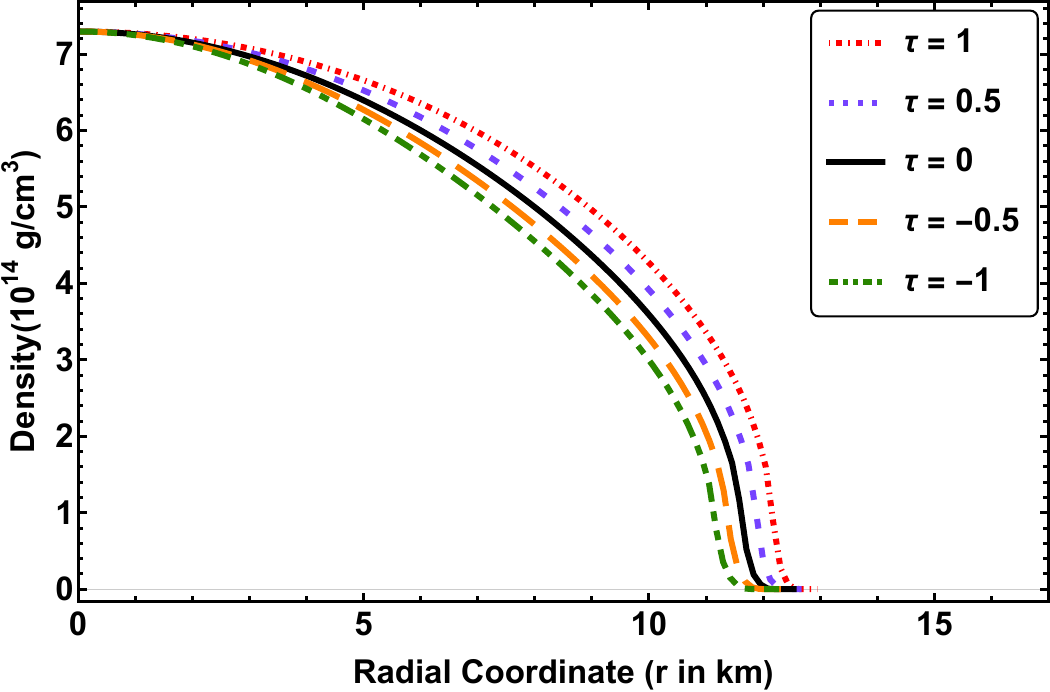}
	\caption{The density profiles inside neutron stars for various values of the anisotropic strength, each for $ \rho_c = 7.2955 \times 10^{14}~{\rm g \, cm^{-3}}$ and BSk21 EoS.}
	\label{density}
\end{figure}

In Fig. \ref{density}, we plot the radial coordinate (in km) inside the neutron star along the horizontal axis and the corresponding density in ${\rm 10^{14}~ g \, cm^{-3}}$ along the vertical axis, for BSk21 EoS and different values of $\tau$, each with $\rho_c = 7.2955 \times 10^{14}~{\rm g \, cm^{-3}}$ (the point where all the curves meet on the vertical axis). We see that for each of the cases, the density consistently decreases as the radial coordinate increases and eventually becomes almost zero at the surface of the star.

\begin{figure}[htbp]
	\centering
	\includegraphics[width=0.5\textwidth]{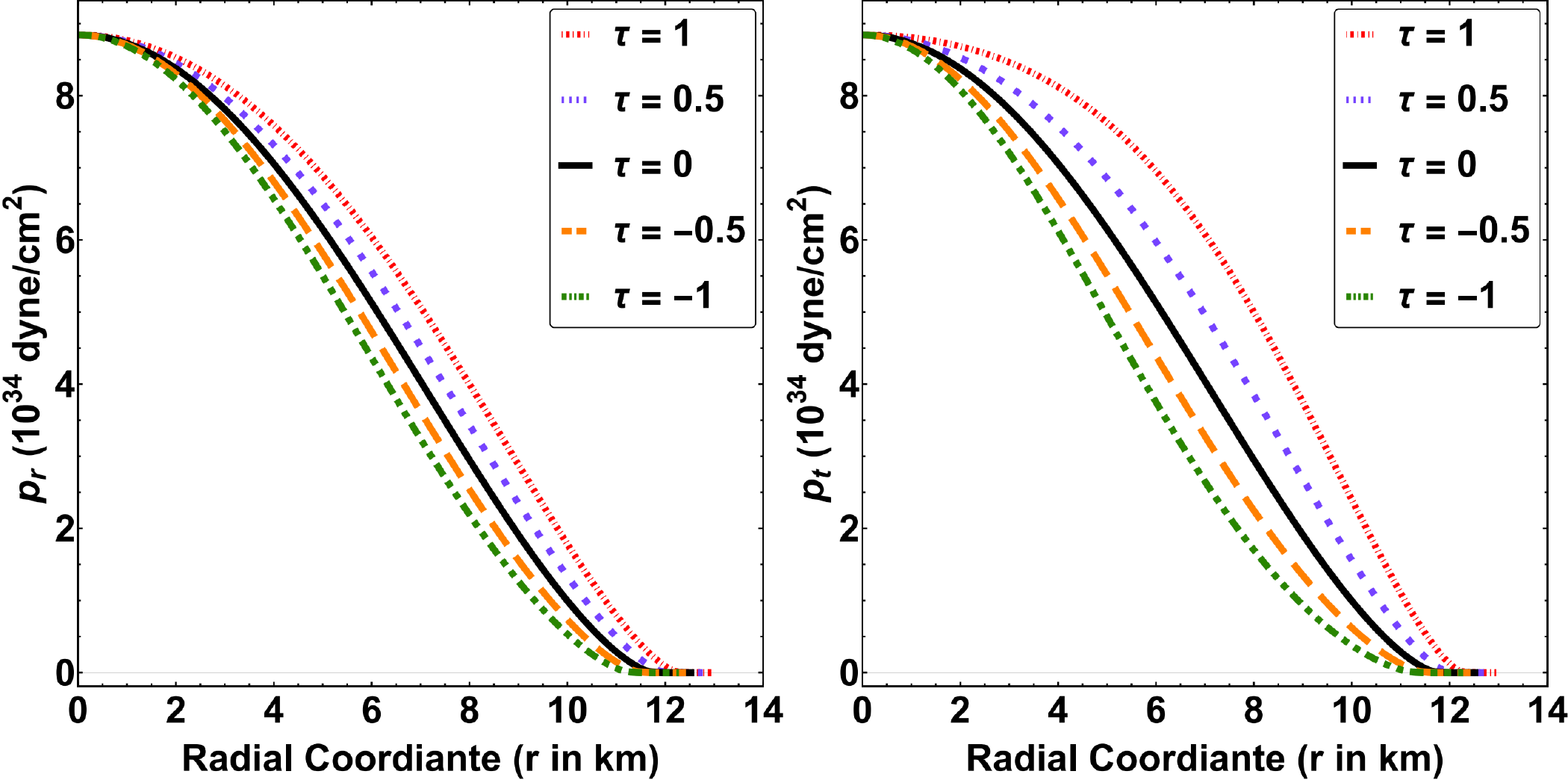}
	\caption{The variation of the radial (left panel) and the tangential (right panel) pressure inside neutron stars for various values of the anisotropic strength, each for $\rho_c=7.2955 \times 10^{14}~{\rm g \, cm^{-3}}$ and BSk21 EoS.}
	\label{prpt}
\end{figure}

In Fig. \ref{prpt}, we plot the radial coordinate (in km) inside the neutron star along the abscissa and the corresponding pressure along the ordinate for BSk21 EoS and different values of $\tau$. The left panel of the figure has the radial pressure along the ordinate while the right panel has the tangential pressure along the ordinate for the same values of $\tau$. Both of the plots are for $\rho_c = 7.2955 \times 10^{14}~{\rm g \, cm^{-3}}$. We see that for all values of $\tau$, both $p_r$ and $p_t$ decreases monotonically with the increase of $r$ and ultimately reaches zero at the surface of the star.

\section{f-mode Oscillations of anisotropic neutron stars}\label{sec:perscheme}
\subsection{Analytical Setup}
\label{subsec:perschemeAnal}
To derive the differential equations governing the f-mode oscillations of neutron stars in general relativity, we have to take into account the perturbations in the metric as well as in the matter in the neutron star. First, we decompose the perturbed metric into two parts, a background and a perturbation on the background. So, the perturbed metric can be written as: 
\begin{equation}
	g_{\mu\nu}  = g_{\mu\nu}^{(B)} + h_{\mu\nu},
\end{equation} 
where $g_{\mu\nu}^{(B)}$ is the background metric for the spherically symmetric static star, described in Eq. (\ref{metric}), and $h_{\mu\nu}$ is the perturbation on it up to the linear order. The linear order perturbation of the static spherically symmetric metric is decomposed into spherical harmonics $Y^m_l(\theta, \phi)$, a function that has a radial dependence, and a function with the time dependence. The perturbed metric is accompanied by a perturbation in the energy-momentum tensor. The coupling between the perturbation of the metric and the perturbation in the energy-momentum tensor is described by the linearized perturbed Einstein equations,
\begin{equation}\label{linein}
	\delta G_{~\alpha}^{\beta} = 8 \pi \delta T_{~\alpha}^{\beta},
\end{equation}
where $\delta G_{~\alpha}^\beta$ is the perturbed Einstein tensor and $\delta T_{~\alpha}^\beta$ is the perturbed energy-momentum tensor. $G_{~\alpha}^\beta$ and $T_{~\alpha}^\beta$ are the un-perturbed Einstein tensor and the energy-momentum tensor, respectively. The expression for the linearized perturbed Einstein tensor, $\delta G_{\mu \nu}$, is given by \cite{Kojima:1992ie}:
\begin{widetext}
\begin{equation}\label{perteinst}
	\begin{split}
		{\delta G_{\mu\nu}} = -\frac{1}{2} [ \nabla^\alpha \nabla_\alpha h_{\mu\nu} - (\nabla_\nu f_\mu + \nabla_\mu f_\nu) + 2 {{{\mathcal{R}^\alpha}_\mu}^\beta}_\nu h_{\alpha\beta} + \nabla_\nu \nabla_\mu{{h^\alpha}_\alpha}
		- ({\mathcal{R}^\alpha}_\nu h_{\mu\alpha} + {\mathcal{R}^\alpha}_\mu h_{\nu\alpha}) +\\ 
		g^{(B)}_{\mu\nu} (\nabla^\alpha f_\alpha + \nabla^\beta \nabla_\beta {h^\alpha}_\alpha) + \mathcal{R} h_{\mu\nu}
		- g^{(B)}_{\mu\nu} \mathcal{R}^{\alpha\beta} h_{\alpha\beta}],
	\end{split}
\end{equation}
\end{widetext}
where $f_\nu = \nabla^\beta h_{\nu\beta}$ and $\mathcal{R}_{\alpha \beta \gamma \delta}$, $\mathcal{R}_{\alpha \beta}$, and $\mathcal{R}$ are Riemann tensor, Ricci tensor, and Ricci scalar, respectively, of the background metric and $\nabla_\sigma$ denotes the covariant derivative along any coordinate $x^\sigma$. In the present work, we restrict ourselves to even parity perturbation (polar modes) for fixed values of $l$ and $m$. In Regge-Wheeler gauge \cite{Regge:1957td}, the even mode of perturbation takes the form:
\begin{widetext}
\begin{equation}\label{pertmet}
	h_{\alpha\beta} = 
	-r^l
	\begin{pmatrix}
		e^\nu  H_0(r)      & i \omega r H_1(r) & 0           & 0                          \\
		i \omega r H_1(r) & e^\lambda H_2(r)  & 0           & 0                          \\
		0                     & 0                     & r^2 K(r) & 0                          \\
		0                     & 0                     & 0           & r^2 \sin^2 \theta K(r)
	\end{pmatrix}
	 Y_l^m (\theta, \phi) e^{i \omega t}~,
\end{equation}
\end{widetext}
where $H_0(r)$, $H_1(r)$, $H_2(r)$, $K(r)$ are metric perturbation functions that have only radial dependence, and $\omega$ is the angular frequency of the time varying component of the perturbation. We can further reduce the independent components of the metric perturbation. If we subtract the $``33"$ component of the Einstein equation from the $``22"$ component, we get the following: 
\begin{subequations}
	\begin{align}
	\delta (G_{~2}^{2} - G_{~3}^{3}) &= 8 \pi  \delta (T_{~2}^{2} - T_{~3}^{3}) \\
	&= 8 \pi (\delta p_t - \delta p_t) \\
	&= 0~,
	\end{align}
\end{subequations}
where we have used the fact that $T_{~\alpha}^\beta = diag(-\rho, p_r, p_t, p_t)$. The left hand side of the equation can easily be calculated from Eq. (\ref{perteinst}). After some algebra, we find that $H_0(r) = H_2(r)$. Hereafter, we will write $H_0(r) = H_2(r) = H(r)$.

After defining the metric perturbation, now we focus on the perturbation of the matter of the neutron star. The perturbation of the fluid in the star is described by the Lagrangian displacement vector $\xi^\alpha$, which has components \cite{Gittins:2020mll, Doneva:2012rd, Detweiler:1985zz} as follows:
\begin{equation}\label{pertfluid}
	\xi^\mu =  
	\begin{pmatrix}
		0                                          \\
		r^{l-1} \, e^{-\lambda/2} W(r)                     \\
		- r^{l-2} \, V(r) {\partial}_\theta               \\
		-\frac{r^{l-2}}{\sin^2 \theta} \, V(r) {\partial}_\phi
	\end{pmatrix}
	Y_l^m (\theta, \phi) e^{i \omega t},
\end{equation}
where $V(r)$ and $W(r)$ are fluid perturbation functions. We will use this Lagrangian displacement vector extensively to describe the perturbation of the energy-momentum tensor. The Lagrangian variation of the four-velocity ($u^\mu$) can be written as \cite{Andersson:2020phh}: 
\begin{equation}
	\Delta u^\mu =  \frac{1}{2} u^\mu u^\nu u^\sigma \Delta g_{\nu\sigma},
\end{equation}
where the Lagrangian perturbation of the metric is: 
\begin{equation}
	\Delta g_{\alpha\beta} = h_{\alpha\beta} + \nabla_\alpha \xi_\beta + \nabla_\beta \xi_\alpha~.
\end{equation}
These Lagrangian perturbations are related to the Eulerian perturbations by the relation 
\begin{equation} \label{eq:LagrangianEulerianreln}
\Delta = \delta + \mathcal{L}_{\xi} ,
\end{equation} 
where $\delta$ denotes the Eulerian variation and $\mathcal{L}_\xi$ is the Lie derivative along $\xi^{\alpha}$. The explicit expression for the Eulerian variation of the velocity-four vector is given by 
\begin{equation}\label{pertvel}
	\delta u^\mu = (\delta^{\mu}_{~\rho} + u^\mu u_\rho)(u^\sigma \nabla_\sigma\xi^\rho - \xi^\sigma \nabla_\sigma u^\rho) +\frac{1}{2} u^\mu u^\sigma u^\rho h_{\sigma\rho}. 
\end{equation}
After substituting the expressions of $u^\sigma$, $\xi^\sigma$ and $h_{\alpha\beta}$ to Eq. (\ref{pertvel}), we get 
\begin{equation}\label{deltau}
	\delta u^\sigma = 
	\begin{pmatrix}
		-\frac{H}{2} \\
		i \omega r^{-1} \, e^{-\lambda/2} W(r) \\
		-i \omega r^{-2} \, V(r) \partial_{\theta} \\
		-i \omega (r \sin \theta)^{-2} \, V(r) \partial_{\phi}
	\end{pmatrix}
	r^l e^{-\nu /2} Y_l^m (\theta, \phi) e^{i \omega t} .
\end{equation}
We also expand the perturbation of the radial unit vector, $s^\mu$, in harmonics as follows:
\begin{equation}\label{deltas}
	\delta s^\sigma =
	\begin{pmatrix}
		i \omega S_0(r)\\
		S_1(r)\\
		0\\
		0
	\end{pmatrix}
	r^l \, Y_l^m (\theta, \phi) \, e^{i \omega t} ,   
\end{equation}
where $S_0(r)$ and $S_1(r)$ are two functions of the radial coordinates, which we need to determine. For the sake of brevity, we will drop the independent variables (coordinates) from the functions for the rest of the paper, i.e., instead of $H(r)$, $H_1(r)$, $K(r)$, $W(r)$, $V(r)$, $S_0(r)$, $S_1(r)$, and $Y_l^m (\theta, \phi) $, we will simply write $H$, $H_1$, $K$, $W$, $V$, $S_0$, $S_1$, and $Y_l^m$, respectively.

As $s^\sigma$ is a space-like unit radial vector, so, up to the linear order, it should satisfy, $\delta s^\sigma s_\sigma  + s^\sigma \delta s_\sigma = 0$, as well as $\delta s^\sigma u_\sigma  +  s^\sigma \delta u_\sigma = 0$. These two conditions allow us to write $S_0$ and $S_1$ as:
\begin{eqnarray}
	S_0 &=& \frac{e^{-\nu}}{r} W - e^{-(\nu + \lambda/2)} \, r \, H_1 ~,\\
	S_1 &=& \frac{e^{-\lambda/2}}{2} H ~.
\end{eqnarray}  
Now, to describe the perturbation in the density and the pressures (both the radial and the tangential), we need to know the perturbation of the number density. The Lagrangian perturbation in the number density of the particles can be derived from the conservation of the number density current, which is given by $\nabla_\alpha n^\alpha = 0$, where $n^\alpha = n u^\alpha$, and $n$ is the number density of the particles\footnote{In some works \cite{Karlovini:2002fc}, anisotropy in the number density has been studied by introducing the concept of the linear number density $n_1$, $n_2$, and $n_3$ such that $n=n_1 \, n_2 \, n_3$, where $n$ is the volume number density. One can even simplify this to $n_1 = n_r$ and $n_2 = n_2 =n_t$, where $n_r$ and $n_t$ are the linear number densities in the radial and the tangential directions, respectively. In the present work, we did not consider any anisotropy in the number density.}. The Lagrangian perturbation in the number density of particles in the neutron star is given by \cite{Andersson:2020phh, Misner:1973prb}:
\begin{equation}
	\Delta n = -\frac{1}{2} n \perp^{\mu\nu} \Delta g_{\mu\nu} ~,
\end{equation} 
where $\perp^{\mu\nu}$ is the projection operator, which is orthogonal to the fluid flow. $\perp^{\mu\nu}$ can be expressed as:
\begin{equation}
	\perp^{\mu\nu} = u^{\mu}u^{\nu} +g^{\mu\nu}.
\end{equation}
Computing explicitly, we can write the expression of $\Delta n$ as,
\begin{equation}\label{pertn}
	\Delta n = -\frac{1}{2} n \perp_g Y^m_l e^{i\omega t},
\end{equation}
where,
\begin{eqnarray} \label{perpg}
	\perp_g  = - 2 r^l  &\Bigg[&\Bigg(K+\frac{1}{2} H \Bigg) - \frac{l(l+1)}{r^2} V \nonumber \\
	&-& \frac{e^{-\lambda/2}}{r}W' - \Bigg(\frac{l+1}{r^2} e^{-\lambda/2}\Bigg) W \Bigg] ,
\end{eqnarray} 
which is similar to the expression described by \citet{Comer:1999rs}.
Now, following \citet{thorne_campolattaro_1967} and \citet{Lindblom:1983ps}, we make some assumpitions: (i) there is no structural anisotropy within the star, and as a result, the perturbation in the Lagrangian strain tensor is zero. This simplification might not be valid when one incorporates the effects of the magnetic field and/or rotations. (ii) Due to the absence of structural anisotropy, the work done in an infinitesimally small volume $dV$ can be taken as $p_{\rm avg} \, dV$ where $p_{\rm avg}$ is the pressure averaged over all directions. (iii) The matter is barotropic, i.e., the density $(\rho)$ is a function of the number density only, i.e., $\rho = \rho(n)$. Under the above approximations, the perturbation in the density can be written as,
\begin{equation}\label{pertrho}
	\Delta \rho  = \frac{d \rho }{d n} \Delta n = \kappa \Delta n,
\end{equation}
where $\kappa$ is the chemical potential $(\kappa = d\rho/d n)$.  Now we can use the Gibbs relation, which is given by \cite{Andersson:2020phh}: 
\begin{equation}\label{gibbs}
\rho + p_{\rm avg} = \kappa n ~,
\end{equation}
where the average pressure $p_{\rm avg}$ can be written as \cite{Rezzolla2017-mt}:
\begin{equation}
	p_{\rm avg} = \frac{1}{3}\left(T_{~1}^{1} + T_{~2}^{2} + T_{~3}^{3} \right) = \frac{p_r + 2 p_t}{3} = p_r + \frac{2}{3}\chi .
	\label{avpress}
\end{equation} Although $p_{\rm avg}$ is a function of the radial coordinate inside the star, following the convention of this paper, we write $p_{\rm avg}$ instead of $p_{\rm avg}(r)$. Note that, there will be extra terms in Eq. (\ref{gibbs}) if structural anisotropy is considered \cite{Carter:1972, Lau:2024oik}.  

Using Eq. (\ref{pertn}) in Eq. (\ref{pertrho}) and the Gibbs relation (Eq. (\ref{gibbs})), we get,
\begin{equation}
	\Delta \rho   = -\frac{1}{2} (p_{\rm avg} + \rho ) \perp_g Y^m_l e^{i \omega t} ~. 
\end{equation}
The Lagrangian variation of the radial pressure can be written as:
\begin{equation}\label{prlag}
	\Delta p_r = \frac{d p_r}{d\rho } \Delta \rho = - \frac{1}{2} \frac{dp_r}{d\rho} ( p_{\rm avg} + \rho ) \perp_g Y^m_l e^{i \omega t} ~.
\end{equation}
Using the definition of the speed of sound in the radial direction $v_{sr}^2=\frac{\partial p_r}{\partial \rho} = \frac{d p_r}{d \rho}$ as mentioned in Sec. \ref{sec:eosAnisotropy}, Eq. (\ref{prlag}) can also be written as:
\begin{equation}\label{prlagNew}
	\Delta p_r =  - \frac{1}{2} v_{sr}^2 ( p_{\rm avg} + \rho ) \perp_g Y^m_l e^{i \omega t} ~.
\end{equation}
From the relation between the Lagrangian and the Eulerian variations, i.e., Eq. (\ref{eq:LagrangianEulerianreln}), we can write the relation between the Lagrangian perturbation and the Eulerian perturbation of the radial pressure as:
\begin{eqnarray}\label{prlageuler}
	\Delta p_r &&= \delta p_r + \xi^r p'_r \nonumber\\
	 &&= \delta p_r + r^{l-1} \, e^{-\lambda/2}\left[-\frac{\nu'}{2}(\rho + p_r)+\frac{2\chi}{r}\right]W \, Y_l^m e^{i \omega t}.\nonumber\\ 
\end{eqnarray} From Eqs. (\ref{prlag}) and (\ref{prlageuler}), we get the expression for the Eulerian variation of the radial pressure, $\delta p_r$. With the help of $\delta p_r$, we calculate the Eulerian variation of $\rho$ and $p_t$, which are given by:
\begin{eqnarray}
	\delta \rho  &=& \frac{d \rho }{d p_r} \delta p_r , \\
	\delta p_t &=& \delta \chi + \delta p_r. 
\end{eqnarray}
 With these expressions in our hand, we are ready to calculate the perturbation of the energy-momentum tensor, namely $\delta T_{~\alpha}^{\beta}$, which can be written as:
\begin{align}\label{enmomper}
	\delta T_{~\alpha}^{\beta} = (\delta \rho  + \delta p_t) u^\beta u_\alpha+ (\rho + p_t) (u^\beta \delta u_\alpha +\delta u^\beta u_\alpha ) \nonumber\\
	+ \delta_{~\alpha}^{\beta} \delta p_t + (\delta p_r - \delta p_t) s^\beta s_\alpha + (p_r - p_t)(s^\beta \delta s_\alpha + \delta s^\beta s_\alpha )~.
\end{align}
Using Eqs. (\ref{deltau}), (\ref{deltas}), and (\ref{enmomper}), as well as $u^\beta$ and $s^\beta$, we can calculate the non-zero components of the perturbed energy-momentum tensor, which are given by:
\begin{subequations}\label{delTmunu}
\begin{align}
\delta T_{~0}^{0} &=  - \delta \rho = -\delta \tilde{\rho}(r) \, r^l Y_l^m e^{i \omega t}~,\\
\delta T_{~0}^{1} &= -i \omega e^{-\lambda/2} r^{-1} [p_r + \rho] W r^l Y_l^m e^{i \omega t}~,\\
\delta T_{~0}^{2} &= i \omega r^{-2} [p_r + \rho + \chi] V r^l ( \partial_\theta Y_l^m ) e^{i \omega t}~,\\
\delta T_{~0}^{3} &= i \omega r^{-2} [p_r + \rho + \chi] V r^l (\sin\theta)^{-2} ( \partial_\phi Y_l^m ) e^{i \omega t}~,\\
\delta T_{~1}^{1} &= \delta p_r = \delta \tilde{p}_r(r) \, r^l Y_l^m e^{i \omega t}~, \label{delradp}\\
\delta T_{~2}^{2} &= \delta p_t = [\delta \tilde{p}_r(r) + \delta \tilde{\chi}(r)] \, r^l Y_l^m e^{i \omega t}~,\\
\delta T_{~3}^{3} &= \delta p_t = [\delta \tilde{p}_r (r) + \delta \tilde{\chi}(r)] \, r^l Y_l^m e^{i \omega t}~,    
\end{align}
\end{subequations}
where $\delta \rho$, $\delta p_r$ and $\delta \chi$ have been decomposed into radial, angular, and temporal parts as, $\delta \rho = r^l \, \delta \tilde{\rho}(r) \, Y_l^m e^{i \omega t}$, $ \delta p_r = r^l\, \delta \tilde{p_r}(r) \, Y_l^m e^{i \omega t}$ and $ \delta \chi = r^l \, \delta \tilde{\chi}(r) \, Y_l^m e^{i \omega t}$. In the above expressions, if we set $\chi = 0$ and $\delta \chi = 0$, the resulting expressions would resemble those of the isotropic case described by \citet{thorne_campolattaro_1967}. 

Now, we use the linearly perturbed Einstein equations (Eqs. (\ref{linein})), and the equation of conservation of the energy-momentum tensor up to the linear order, i.e., $\delta(\nabla_\nu T^{\mu\nu}) = 0$, to derive equations of oscillation for the metric perturbation variables ($H$, $H_1$, $K$) as well as the fluid perturbation variables ($V$ and $W$). First, we perform a change of variable, which simplifies the boundary condition. We know that, at the surface of the star $(r = R)$, the Lagrangian perturbation of the radial pressure is zero, i.e., $\Delta p_r = 0$. We introduce a new variable $X$ to write $\Delta p_r$ in the form  
\begin{equation}\label{xdef}
	\Delta p_r =  - r^l e^{-\nu /2} X Y_l^m e^{i \omega t}. 
\end{equation}
This change of variable allows us to write one of the perturbation equations in terms of $X$, and we can impose the boundary condition $X = 0$ at $r=R$, when solving the oscillation equations numerically. Using the expressions from equations (\ref{perpg}), (\ref{prlagNew}), and (\ref{xdef}), we extract the expression of $X$, which is given by:
\begin{eqnarray}\label{Xnew}
	X = (\rho &+& p_{\rm avg} )v_{sr}^2\Big[\frac{e^{\nu/2 - \lambda/2}}{r} W' + \frac{(l+1) e^{\nu/2 - \lambda/2}}{r^2} W \nonumber \\
	&&+ \frac{l(l+1) e^{\nu/2}}{r^2} V - e^{\nu/2} K - \frac{e^{\nu/2}}{2} H\Big] .
\end{eqnarray}
Using Eqs. (\ref{prlageuler}) and (\ref{xdef}), we can write the Eulerian perturbation of the radial pressure as:
 \begin{align}\label{delpr}
 	\delta p_r  = \delta \tilde{p}_r r^l Y_l^m e^{i \omega t} = \bigg[& - e^{-\nu/2} X + \frac{W}{2 r^2}e^{-\lambda/2} \biggl\{ - 4 \chi \nonumber\\ 
 	&+ r \nu' (p_r + \rho) \biggr\}  \bigg] r^l Y_l^m e^{i \omega t}~.
 \end{align}
It is obvious that $X$ depends only on the radial coordinate $r$, and following the convention of this paper, we write it simply as $X$. Using this new variable $X$, we write the governing equations of oscillations as:
\begin{widetext}
	\begin{align}
		H_1' &= \frac{e^\lambda}{r} H + \frac{e^\lambda}{r} K - \frac{ (l+1) +2 m r^{-1} e^\lambda + 4 \pi r^2 e^\lambda (p_r - \rho)}{r} H_1 - \frac{16 \pi e^\lambda (\rho + p_r + \chi)}{r}V , \label{eq:H1}\\
		K' &= \frac{1}{r} H  -\left[(l+1)r^{-1} - \frac{1}{2}\nu'\right]K + \frac{l(l+1)}{2 r}H_1 - \frac{8 \pi e^{\lambda/2}(p_r + \rho)}{r} W ,\label{eq:K}\\
		W' &= \frac{1}{2}  e^{\lambda /2} r H + e^{\lambda /2} r K-\frac{e^{\lambda /2} l (l+1)}{r} V-\frac{(l+1) }{r} W+\frac{r e^{\frac{\lambda -\nu }{2}}}{v_{sr}^2 \left(p_r+\rho +\frac{2\chi }{3}\right)} X ,\label{eq:W}\\
		X' &= -2\frac{e^{\nu/2}}{r}\delta\tilde{\chi} - \left[\frac{l}{r} + \frac{\chi (6 + r \nu')}{r v_{sr}^2 (3(p_r + \rho ) + 2 \chi)}\right] X + (p_r + \rho) e^{\nu/2}\left[\frac{1}{2 r} - \frac{\nu'}{4}\right] H \nonumber \\
		&\quad-\left[\frac{ e^{\nu/2}l(l+1)(p_r+\rho)\nu'}{2 r^2} - \chi \frac{2 l(l+1) e^{\nu/2}}{r^3} \right] V \nonumber\\
		&\quad+ \left[\frac{e^{\nu/2}(p_r + \rho)}{2}\left(\frac{l(l+1)}{ 2 r } + r \omega^2 e^{-\nu}\right) + \frac{\chi e^{\nu/2}l(l+1)}{2r}\right] H_1 \nonumber \\
		&\quad+ \left[\frac{1}{2} e^{\nu/2}(p_r + \rho) \left(\frac{3}{2}\nu' -\frac{1}{r}\right) + \chi \frac{e^{\nu/2}(-6 + r \nu')}{2 r }\right] K \nonumber \\
		&\quad+ \left[-\frac{e^{\nu/2} (p_r +\rho)}{r}\left(e^{\lambda/2 - \nu}\omega^2 + 4 \pi e^{\lambda/2}(p_r + \rho) - \frac{1}{2} r^2 (r^{-2} e^{-\lambda/2} \nu')'\right) \right. \nonumber \\
		&\quad\left.- \chi' \frac{2 e^{(\nu - \lambda)/2}}{r^2}+ \frac{e^{(\nu + \lambda)/2}}{r^3}\chi \left(6 - 14 m r^{-1} - 8 \pi r^2 p_r\right)\right] W , \label{eq:X}
	\end{align}
\end{widetext}
where $H_1$ and $K$ were introduced in Eq. (\ref{pertmet}), $W$ in Eq. (\ref{pertfluid}), and $X$ in Eq. (\ref{xdef}). The remaining functions in Eq. (\ref{pertmet}) and (\ref{pertfluid}), namely $H_0 = H_2 = H$ and $V$ are related to other functions by following algebraic relations:
\begin{widetext}
	\begin{equation}
		\begin{aligned}
			\Biggl[3 m &+ \frac{1}{2} (l - 1) (l + 2) r+ 4 \pi r^3 p_r \Biggr] H = 8 \pi e^{-\nu/2} r^3 X - \Biggl[- e^{-\nu-\lambda} r^3 \omega^2 + \frac{1}{2}l (l+1) (m + 4 \pi r^3 p_r)\Biggr]H_1 \\
			&\quad + \chi 16 e^{-\lambda/2} \pi r W - \Biggl[-\frac{1}{2}(l-1)(l+2) r + e^{-\nu} r^3 \omega^2 + \frac{e^{\lambda}}{r} (m + 4 \pi r^3 p_r)(3m -r + 4\pi r^3 p_r)\Biggr]K, \label{eq:H}
		\end{aligned}
	\end{equation}
	
	\begin{equation}
		\omega^2 (\rho + p_r + \chi) V = e^{\nu/2} X - \frac{1}{2}e^\nu (p_r + \rho) H + \frac{e^{\nu - \lambda/2} p'_r}{r} W - e^\nu \delta \tilde{\chi}~. \label{eq:V}
	\end{equation}
\end{widetext}

These are the set of equations that govern the oscillations of anisotropic neutron stars in general relativity. The equations for the metric perturbation dynamics, i.e., the equation for $H_1'$ has been obtained from $\delta G_{~0}^2 = 8 \pi \delta T_{~0}^2$ and the equation for $K'$ has been obtained from $\delta G_{~0}^1 = 8 \pi \delta T_{~0}^1$. The expressions for $\delta G_{~\nu}^{\mu}$ are obtained by putting $h_{\mu \nu}$ from Eq. (\ref{pertmet}) in Eq. (\ref{perteinst}) and raising one index. The explicit expressions for $\delta T_{~\mu}^{\nu}$ is given in Eq. (\ref{delTmunu}). On the other hand, the dynamics of the fluid perturbations, which are given by $W'$ and $X'$, have been obtained with the help of Eq. (\ref{Xnew}) and $\delta(\nabla_\mu T_{~\, 1}^{\mu}) = 0$, respectively. Eq. (\ref{eq:H}), which relates the metric perturbation variable $H$ to other variables, has been obtained by combining $\delta G_{~1}^1 = 8 \pi \delta T_{~1}^1$ and $\delta G_{~1}^2 = 8 \pi \delta T_{~1}^2$. Eq. (\ref{eq:V}) relates the fluid perturbation variable $V$ to other variables and has been obtained from $\delta(\nabla_\mu T_{~\, 2}^{\mu}) = 0$. The effect of anisotropy is manifested in these equations through terms having $\chi$, $\chi'$, and $\delta \chi$. These terms are zero in the isotropic case and our equations take the well known forms given by \citet{Detweiler:1985zz}.
 
For the choice of our ansatz of anisotropy, we get
\begin{equation}\label{delchiansatz}
	\delta \chi = \frac{\partial \chi}{\partial p_r} \delta p_r + \frac{\partial \chi}{\partial \mu} \delta \mu ~.
\end{equation}
Note that, unlike Cowling approximation, in our full general relativistic framework, the variable $\mu$ is perturbed. The quantity $\mu = 2 m /r = 1 - e^{-\lambda}$, which describes the local compactness of the star, can be perturbed as $\delta \mu = e^{-\lambda} \delta \lambda$, where $\delta\lambda$ represents the perturbation of $\lambda$. As we are interested up to the linear order perturbation, so from the perturbation of the metric, we can write
\begin{equation}\label{expdellamb}
	e^{\lambda + \delta \lambda} = e^\lambda(1 - H r^l Y_l^m e^{i \omega t})~.
\end{equation}
Neglecting the higher order perturbation terms (O$((\delta \lambda)^2)\dots$), we can write
\begin{equation}\label{dellambda}
	\delta \lambda  = - H r^l Y_l^m e^{i \omega t}~.
\end{equation}
Using Eqs. (\ref{quasiparam}), (\ref{delradp}), and (\ref{dellambda}) in Eq. (\ref{delchiansatz}), we get 
\begin{equation}
	\delta \chi  = \tau \left[\left(\frac{2m}{r}\right)\delta \tilde{p}_r - \left( p_r\left(1 - \frac{2 m}{r}\right) H \right)\right] r^l Y_l^m e^{i \omega t}~,
\end{equation} 
where $\delta \tilde{p}_r$ is given in equation (\ref{delpr}).

\subsection{Numerical Techniques}  
\label{subsec:perschemeNum}            
Equipped with the above set of equations, we are now prepared to solve them numerically. Since our primary objective is to determine the quasi-normal modes, it is important to carefully consider the specific initial value and boundary value conditions. In contrast to the background case, here we are confronted with the task of handling seven first-order differential equations (three for the background and four for the perturbations) as well as two simultaneous algebraic relations. It is crucial to ensure that the equations remain regular at the center of the star, while the boundary condition dictates that the Lagrangian perturbation of the radial pressure $(\Delta p_r)$ must be zero at the outermost surface of the perturbed star. By examining the set of equations, we observe that they exhibit singularity at $r = 0$. 

To ensure regularity at the center, we expand all the variables at a point $r=R/10^6$ as a Taylor series around $r=0$. The variables for which we do such expansions include the variables that describe the unperturbed star, e.g., $\lambda$, $\nu$, $m$, $\rho$, $p_r$, $\chi$ as well as the perturbation variables $H$, $H_1$, $K$, $V$, $W$, and $X$. We substitute these expansions into the equations that govern the unperturbed star (Eqs. (\ref{eq2}), (\ref{inner}), (\ref{eq:modTOV})), as well as perturbation equations (Eqs. (\ref{eq:H1}), (\ref{eq:K}), (\ref{eq:W}), (\ref{eq:X}), (\ref{eq:H}), and (\ref{eq:V})) to derive the necessary conditions that are:
\begin{align}
	H_1(0) &= \frac{1}{l(l+1)} \left[ 2 l K(0) +  16 \pi \left\lbrace \rho(0) + p_r(0) \right\rbrace W(0) \right],\label{H10}\\
	X(0)   &= \left\lbrace  p_r(0) + \rho(0) \right\rbrace \, e^{\nu(0)/2}  \left[ \left\lbrace \frac{4\pi}{3}\left\lbrace \rho(0)+ 3 p_r(0) \right\rbrace \right. \right. \nonumber\\
	& \quad \left. \left. - \frac{\omega^2 e^{-\nu(0)}}{l} \right\rbrace W(0)+ \frac{1}{2} K(0) \right] - e^{\nu(0)/2} \chi_2(0) W(0),\label{X0}
\end{align} 
where $\chi_2(0)$ is the second order in the expansion of $\chi(r)$. Since the radial and the tangential pressures are equal at the center, we have $\chi(0) = \delta\chi(0) = 0$. This condition guarantees that at the center, the radial pressure is equal to the tangential pressure, and the Eulerian perturbation of the radial pressure is equal to the Eulerian perturbation of the tangential pressure. Based on our ansatz (Eq.(\ref{quasiparam})), we can express $\chi_2(0)$ as $\chi_2(0) = \tau \frac{16 \pi}{3} \rho(0) p_r(0)$.
From Eqs. (\ref{H10}) and (\ref{X0}), we see that $H_1(0)$ and $X(0)$ depend on two independent variables $K(0)$ and $W(0)$. These expressions exhibit similarity to those of the isotropic case described by \citet{Detweiler:1985zz}, with the addition of the term containing $\chi_2(0)$ that arises due to the presence of the anisotropy. Now, we need to set the values of $W(0)$ and $K(0)$ in such a way that after integrating Eqs. (\ref{eq:H1}), (\ref{eq:K}), (\ref{eq:W}), and (\ref{eq:X}) along with Eqs. (\ref{eq2}), (\ref{inner}), (\ref{eq:modTOV}), radially outward from the point $R/10^6$, we would get $X(R)=0$. To do this, following \citet{Detweiler:1985zz}, we use two sets of values of $W(0)$ and $K(0)$, namely, $W(0) = 1$, $K(0) = + [ \rho(0)+p_r(0)]$ and $W(0) = 1$, $K(0) = - [\rho(0)+p_r(0)]$. We perform outward integration with each of the sets independently and get two independent values of $X$ at $r = R$, say $X_{+}$ and $X_{-}$. Then, we combine these two solutions as $X(R) = c_{+} X_{+} + c_{-} X_{-} $ and choose the values of the coefficients $c_{+}$ and $c_{-}$ to ensure $X(R) =0$. For the integration process, we have employed the `LSODA' \cite{doi:10.1137/0904010,Radhakrishnan1993DescriptionAU} method of integration, which dynamically switches between the BDF (Backward Differentiation Formula) method and the Adams method based on the stiffness of the coupled equations. In our implementation, we have set both the relative and the absolute tolerances to $10^{-10}$. This completes the discussion on the integration inside the neutron star.

Outside the star, all the fluid perturbation variables become zero but the metric perturbation variables remain non-zero. This results in a reduction of the number of the equations to two, specifically, the equations for $H_1'$ and $K'$. Following the standard technique, we perform another transformation of variables given by
\begin{align}
	r^l K &= \frac{n_l (n_l +1) r^2 + 3 n_l M r + 6 M^2}{r^2 (n_l r +3 M)} Z + \frac{d Z}{d r^*} , \label{zer1}\\
	r^{l+1} H_1 &= \frac{n_l r^2 - 3 n_l M r - 3 M^2}{(r - 2 M)(n_l r +3 M)}Z + \frac{r^2}{(r - 2 M)}\frac{d Z}{d r^*} ,\label{zer2}
\end{align} 
where $n_l = [l(l+1) /2] - 1$, and $r^* = r + 2 M \,\ln[(r/ 2 M) - 1]$. With these transformations, the differential equations for $H_1'$ and $K'$ can be combined into a single second order differential equation as:
\begin{equation}\label{zerilli}
	\frac{d^2 }{d r^{*2}} \left\lbrace Z(r^*) \right \rbrace \, + \left[ \omega^2  - V_z(r^*) \right] Z(r^*) = 0 ,  
\end{equation} 
where,
\begin{align}
	V_z(r^*) = \frac{(1 - 2M/r)}{r^3 (n_l r + 3 M)^2} \left[ 2 n_l^2 (n_l + 1) r^3 + 6 n_l^2 M r^2 \right. \nonumber \\
	 \left. \quad + 18 n_l M^2 r + 18 M^3 \right] .
\end{align}
Equation (\ref{zerilli}), known as the Zerilli equation, was first derived by \citet{Zerilli:1970se} for the Schwarzschild geometry. The variable $Z(r^*)$ is known as the Zerilli function, which we usually write as $Z$. This method was later used by \citet{Lindblom:1983ps} for the first time for neutron stars. 

To determine the quasi-normal modes, it is necessary to identify the frequencies for which the Zerilli equation satisfies the condition for a purely outgoing wave at the spatial infinity. For this purpose, one needs to numerically integrate the Zerilli equation from the surface of the star ($r=R$) to spatial infinity. In our numerical implementation, we have performed the integration up to a distance of $50~\omega^{-1}$, as the values of $Z$ and $dZ/dr^*$ take the asymptotic values at that distance \cite{Lindblom:1983ps, Zhao:2022tcw}. Here $\omega$ is the angular frequency of the oscillation as introduced in Eq. (\ref{pertmet}). In the large $r$ limit, the solution of the Zerilli equation can be assumed to be purely sinusoidal, which can be interpreted as a composition of the ingoing and the outgoing waves. Denoting the ingoing wave as $Z_{in}(r^*)$ and the outgoing wave as $Z_{out}(r^*)$, we can write the solution at in the large $r$ limit as: 
\begin{equation}
	Z(r^*) = B_{in} Z_{in}(r^*) + B_{out} Z_{out}(r^*) ,
\end{equation} 
where $B_{in}$ and $B_{out}$ are the coefficients of $Z_{in}$ and $Z_{out}$, respectively, which determine the proportions of the in-going and the outgoing waves in the large $r$ limit and are complex conjugate to each other. In this limit, $Z_{in}(r^*)$ and $Z_{out}(r^*)$ can be assumed as power series, which can be written as:
\begin{align}
	Z_{out}(r^*) &= e^{(-i \omega r^*)} \sum_{j = 0}^{\infty} \beta_j r^{-j}~, \label{zin} \\ 
	Z_{in}(r^*) &=  e^{(i \omega r^*)} \sum_{j = 0}^{\infty} \bar{\beta}_j r^{-j}~,\label{zout}
\end{align}
where $\beta_j$s are the coefficients of the power series expansions, and $\bar{\beta}_j$s are the complex conjugates of $\beta_j$s. The coefficients $\beta_j$s can be obtained through a recursion relation by substituting equations (\ref{zin}) and (\ref{zout}) into equation (\ref{zerilli}). To determine the asymptotic behavior of $Z(r^*)$, we consider the expansion up to $O(r^{-2})$. The relevant expressions are as follows:
\begin{align}
	\beta_1 &= -i \omega^{-1} (n_l + 1) \beta_0 ~, \\
	\beta_2 &= - \frac{1}{2 \omega^2} \left[ n_l(n_l + 1) - 3 i M \omega\left(1 + \frac{2}{n_l}\right) \right]\beta_0~, 
\end{align}
where $\beta_0$ is an arbitrary constant. For numerical implementation, we have considered $\beta_0  = p_r(0)/\rho(0)$, as suggested by Chandrasekhar and Ferrari \cite{Chandrasekhar:1991fi}. This choice provides a convenient normalization for the coefficients $\beta_j$ in the recursion relation. From these expressions, we can observe that for a fixed mass of the neutron star and for a specific angular mode ($l = 2$ in the present work), the coefficients $B_{in}$ and $B_{out}$ depend solely on the angular frequency $\omega$. To determine the numerical values of $B_{in}$ and $B_{out}$, we can compare the numerically integrated values of the Zerilli function and its first derivative with the corresponding asymptotic analytical values described above. By calculating $B_{in}$ for various frequencies, we can treat it as a function of $\omega$. Consequently, we can find the root of this function, which implies $B_{in}(\omega) = 0$, i.e., purely outgoing wave for that particular $\omega$. The frequency of the oscillation is given by $\mathcal{F} = {\rm Re}(\omega) / 2 \pi$ where ${\rm Re}(\omega)$ is the real part of the root. The inverse of the imaginary part of the root ($\rm 1/ Im(\omega)$) is the damping time of the oscillation. The values of the mode frequency and corresponding damping time obtained by our code for $\tau = 0$ agrees with earlier results reported by \citet{Lu:2011zzd} and \citet{Kunjipurayil:2022zah}. 

\section{Results for f-mode oscillations of anisotropic neutron stars}\label{sec:results}
In the preceding sections, we detailed the analytical and numerical methods used to determine the quasi-normal modes of anisotropic neutron stars. In this section, we aim to quantify the manifestations of these oscillations. As previously mentioned, we primarily use the BSk21 equation of state (EoS) to solve the equations for the unperturbed stellar configuration and its oscillations. In some cases, we also compare the results with those obtained using the softer BSk19 EoS to understand how the softness of the EoS influences the behavior of f-mode oscillations in anisotropic neutron stars.

\begin{figure}
	\centering
	\includegraphics[width=0.5\textwidth]{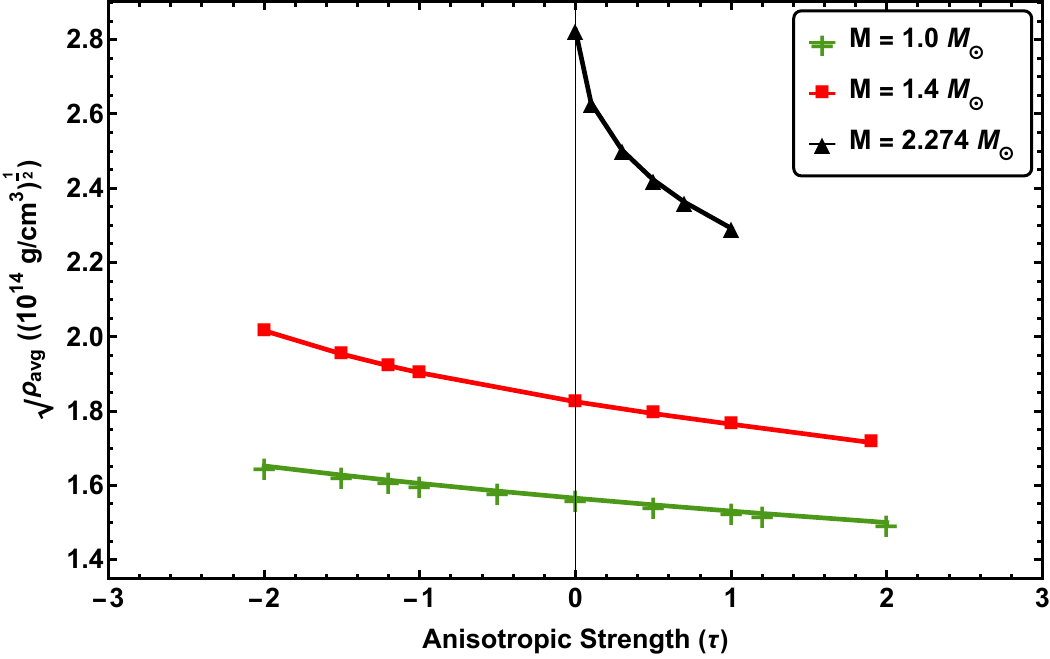}
	\caption{The change in the square root of the average density with the anisotropic strength for various masses of neutron stars for BSk21 EoS.}
	\label{sqrtrhovstau}
\end{figure}

However, to achieve these, first we explore one additional aspect related to the structure of anisotropic stars. We present various physical parameters in Table-\ref{table1}. From this table as well as from Figs. \ref{MR} and \ref{MRbsk19}, it is clear that for any fixed value of the anisotropic strength, at smaller values of the central density ($\rho_c$), an increase in the central density leads to simultaneous increases in the mass ($M$) and the radius ($R$) of the star, up to a certain threshold. In this region, the average density of the stars increases slowly, where the average density of the neutron star is given by $\rho_{\rm avg} = M[(4/3) \pi R^3]^{-1}$. As an example, for the $\tau = 0$ case, this trend holds true up to about $\rho_c = 8.2817 \times 10^{14}~ {\rm g \, cm^{-3}}$, which gives the mass of the star as $1.6~{\rm M_\odot}$. However, above this threshold value of the central density, for each value of the anisotropic strength, an increase in the value of the central density is accompanied by a decrease in the radius and an increase in the mass, resulting in a rapid growth in the average density. To demonstrate the manifestations of this phenomenon, in Fig. \ref{sqrtrhovstau}, we plot the values of the anisotropic strength along the horizontal axis and the values of the square root of the average density ($\sqrt{\rho_{\rm avg}}$) along the vertical axis for neutron stars of masses $1.0~{\rm M_{\odot}}$, $1.4~{\rm M_{\odot}}$, and $2.274~{\rm M_{\odot}}$. These masses represent a possible low mass for a neutron star, the commonly expected mass for a neutron star, and a possible high mass for a neutron star. Moreover, the chosen high value of the mass ($2.274~{\rm M_\odot}$) is the maximum mass of a stable isotropic neutron star for this EoS. Stable neutron stars of mass $1~\rm M_\odot$ exist in the range of $-2 \leq \tau \leq 2$, of mass $1.4 ~\rm M_\odot$ exist in the range of $-2 \leq \tau \leq 1.9$, and of mass $2.274 ~\rm M_\odot$ exist in the range of $0 \leq \tau \leq 1$. From Fig. \ref{sqrtrhovstau}, we see that if the value of $\tau$ increases from 0 to 1, the value of $\sqrt{\rho_{\rm avg}}$ decreases by $2.21 \%$, $3.30 \%$, and $18.86 \%$ for the neutron stars of masses $1.0~{\rm M_{\odot}}$, $1.4~{\rm M_{\odot}}$, and $2.274~{\rm M_{\odot}}$, respectively.

\begin{table*}
	\caption{\label{table1}Numerical values of various parameters for the BSk21 EoS. The columns from left to right contain the values of the anisotropic strength ($\tau$), the central density ($\rho_c$), the mass of the star ($M$), the radius ($R$), the average density ($\rho_{\rm avg}$), the f-mode frequency ($\mathcal{F}$), and the damping time.}
	\begin{ruledtabular}
		\begin{tabular}{lllllll}
			\multicolumn{1}{c}{\begin{tabular}[c]{@{}c@{}}Anisotropic Strength \\ ~ \\ ~ \end{tabular}} & \multicolumn{1}{c}{\begin{tabular}[c]{@{}c@{}}Central Density\\ \\ $(\rm 10^{14}~g \, cm^{-3})$\end{tabular}} & {\begin{tabular}[c]{@{}c@{}} Mass \\ \\ $\rm (M_\odot)$ \end{tabular}}  & {\begin{tabular}[c]{@{}c@{}} Radius \\ \\ (km) \end{tabular}} & \multicolumn{1}{c}{\begin{tabular}[c]{@{}c@{}}Average Density\\ \\ $(\rm 10^{14}~g \, cm^{-3})$\end{tabular}} &  \multicolumn{1}{c}{\begin{tabular}[c]{@{}c@{}}f-mode\\ Frequency\\ (kHz)\end{tabular}} & \multicolumn{1}{c}{\begin{tabular}[c]{@{}c@{}}Damping Time \\ \\ (ms)\end{tabular}} \\
			\hline
			\multicolumn{1}{c}{0}                                                                       & \multicolumn{1}{c}{5.7661}                                                                       & \multicolumn{1}{c}{1.000}              &\multicolumn{1}{c}{12.466}       
			&\multicolumn{1}{c}{2.4501}                                                                                                                                    &\multicolumn{1}{c}{1.564}                                                   &\multicolumn{1}{c}{430.098}                                                   \\
			\multicolumn{1}{c}{0}
			
			&\multicolumn{1}{c}{6.4801}                                                                                                &\multicolumn{1}{c}{1.200}                                     &\multicolumn{1}{c}{12.545}                                   &\multicolumn{1}{c}{2.8846}                                                                                                                                                                                        &\multicolumn{1}{c}{1.638}                                                                               &\multicolumn{1}{c}{313.507}                                                                                \\
			\multicolumn{1}{c}{0}
			&\multicolumn{1}{c}{7.2955}                                                                                                  &\multicolumn{1}{c}{1.400}                                     &\multicolumn{1}{c}{12.589}                                   &\multicolumn{1}{c}{3.3307}                                                                                                                                                                                        &\multicolumn{1}{c}{1.713}                                                                               &\multicolumn{1}{c}{243.028}                                                         						  \\
			\multicolumn{1}{c}{0}
			&\multicolumn{1}{c}{8.2817}
			&\multicolumn{1}{c}{1.600}
			&\multicolumn{1}{c}{12.580}
			&\multicolumn{1}{c}{3.8148}
			&\multicolumn{1}{c}{1.790}
			&\multicolumn{1}{c}{197.672}    	
													\\
			\multicolumn{1}{c}{0}
			&\multicolumn{1}{c}{9.5711}
			&\multicolumn{1}{c}{1.800}
			&\multicolumn{1}{c}{12.496}
			&\multicolumn{1}{c}{4.3779}
			&\multicolumn{1}{c}{1.874}
			&\multicolumn{1}{c}{167.387}    																				\\
			\multicolumn{1}{c}{0}
			&\multicolumn{1}{c}{11.4921}
			&\multicolumn{1}{c}{2.000}
			&\multicolumn{1}{c}{12.294}
			&\multicolumn{1}{c}{5.1085}
			&\multicolumn{1}{c}{1.975}
			&\multicolumn{1}{c}{149.810} 	
			                  						\\
			\multicolumn{1}{c}{0}
			&\multicolumn{1}{c}{22.6476}
			&\multicolumn{1}{c}{2.274}
			&\multicolumn{1}{c}{11.059}
			&\multicolumn{1}{c}{7.9803}
			&\multicolumn{1}{c}{2.317}
			&\multicolumn{1}{c}{154.280} 
			
													\\
			\hline
			\multicolumn{1}{c}{0.5}
			&\multicolumn{1}{c}{5.5206}
			&\multicolumn{1}{c}{1.000}
			&\multicolumn{1}{c}{12.562}
			&\multicolumn{1}{c}{2.3944}
			&\multicolumn{1}{c}{1.585}
			&\multicolumn{1}{c}{396.740}
			\\
			\multicolumn{1}{c}{0.5}
			&\multicolumn{1}{c}{6.1758}
			&\multicolumn{1}{c}{1.215}
			&\multicolumn{1}{c}{12.671}
			&\multicolumn{1}{c}{2.8365}
			&\multicolumn{1}{c}{1.668}
			&\multicolumn{1}{c}{278.210}
			\\
			\multicolumn{1}{c}{0.5}
			&\multicolumn{1}{c}{6.7874}
			&\multicolumn{1}{c}{1.400}
			&\multicolumn{1}{c}{12.737}
			&\multicolumn{1}{c}{3.2155}
			&\multicolumn{1}{c}{1.737}
			&\multicolumn{1}{c}{217.250}
			\\
			\multicolumn{1}{c}{0.5}
			&\multicolumn{1}{c}{8.0285}
			&\multicolumn{1}{c}{1.711}
			&\multicolumn{1}{c}{12.762}
			&\multicolumn{1}{c}{3.9071}
			&\multicolumn{1}{c}{1.858}
			&\multicolumn{1}{c}{155.940}
			\\
			\multicolumn{1}{c}{0.5}
			&\multicolumn{1}{c}{9.1689}
			&\multicolumn{1}{c}{1.925}
			&\multicolumn{1}{c}{12.689}
			&\multicolumn{1}{c}{4.4731}
			&\multicolumn{1}{c}{1.949}
			&\multicolumn{1}{c}{130.850}
			\\
			\multicolumn{1}{c}{0.5}
			&\multicolumn{1}{c}{12.4966}
			&\multicolumn{1}{c}{2.274}
			&\multicolumn{1}{c}{12.252}
			&\multicolumn{1}{c}{5.8688}
			&\multicolumn{1}{c}{2.146}
			&\multicolumn{1}{c}{108.880}
			
			\\

			\multicolumn{1}{c}{0.5}
			&\multicolumn{1}{c}{20.1137}
			&\multicolumn{1}{c}{2.436}
			&\multicolumn{1}{c}{11.209}
			&\multicolumn{1}{c}{8.2100}
			&\multicolumn{1}{c}{2.421}
			&\multicolumn{1}{c}{114.960}
			
			\\
			\hline
			
			\multicolumn{1}{c}{1}
			&\multicolumn{1}{c}{5.3021}
			&\multicolumn{1}{c}{1.000}
			&\multicolumn{1}{c}{12.654}
			&\multicolumn{1}{c}{2.3427}
			&\multicolumn{1}{c}{1.601}
			&\multicolumn{1}{c}{370.810}
			
													\\
			\multicolumn{1}{c}{1}
			&\multicolumn{1}{c}{5.8204}
			&\multicolumn{1}{c}{1.200}
			&\multicolumn{1}{c}{12.776}
			&\multicolumn{1}{c}{2.7310}
			&\multicolumn{1}{c}{1.679}
			&\multicolumn{1}{c}{263.490}
													\\
			\multicolumn{1}{c}{1}
			&\multicolumn{1}{c}{6.3670}
			&\multicolumn{1}{c}{1.400}
			&\multicolumn{1}{c}{12.874}
			&\multicolumn{1}{c}{3.1139}
			&\multicolumn{1}{c}{1.754}
			&\multicolumn{1}{c}{198.690}																					\\
			
			\multicolumn{1}{c}{1}
			&\multicolumn{1}{c}{7.6519}
			&\multicolumn{1}{c}{1.800}
			&\multicolumn{1}{c}{12.955}
			&\multicolumn{1}{c}{3.9296}
			&\multicolumn{1}{c}{1.905}
			&\multicolumn{1}{c}{128.780}
													\\
			\multicolumn{1}{c}{1}
			&\multicolumn{1}{c}{10.0875}
			&\multicolumn{1}{c}{2.274}
			&\multicolumn{1}{c}{12.712}
			&\multicolumn{1}{c}{5.2539}
			&\multicolumn{1}{c}{2.115}
			&\multicolumn{1}{c}{91.680}
													\\
			\multicolumn{1}{c}{1}
			&\multicolumn{1}{c}{14.0191}
			&\multicolumn{1}{c}{2.552}
			&\multicolumn{1}{c}{12.003}
			&\multicolumn{1}{c}{7.0065}
			&\multicolumn{1}{c}{2.335}
			&\multicolumn{1}{c}{85.390}
													\\
			\hline											
			
			\multicolumn{1}{c}{-0.5}
			&\multicolumn{1}{c}{6.0447}
			&\multicolumn{1}{c}{1.000}
			&\multicolumn{1}{c}{12.365}
			&\multicolumn{1}{c}{2.5104}
			&\multicolumn{1}{c}{1.537}
			&\multicolumn{1}{c}{474.480}
			\\
			\multicolumn{1}{c}{-0.5}
			&\multicolumn{1}{c}{6.9786}
			&\multicolumn{1}{c}{1.217}
			&\multicolumn{1}{c}{12.421}
			&\multicolumn{1}{c}{3.0149}
			&\multicolumn{1}{c}{1.613}
            &\multicolumn{1}{c}{345.960}
            \\
			\multicolumn{1}{c}{-0.5}
			&\multicolumn{1}{c}{7.9262}
			&\multicolumn{1}{c}{1.400}
			&\multicolumn{1}{c}{12.425}
			&\multicolumn{1}{c}{3.4638}
			&\multicolumn{1}{c}{1.677}
			&\multicolumn{1}{c}{281.400}
			\\            
			\multicolumn{1}{c}{-0.5}
			&\multicolumn{1}{c}{11.4870}
			&\multicolumn{1}{c}{1.819}
			&\multicolumn{1}{c}{12.170}
			&\multicolumn{1}{c}{4.7917}
			&\multicolumn{1}{c}{1.847}
			&\multicolumn{1}{c}{208.400}			
			\\
			\multicolumn{1}{c}{-0.5}
			&\multicolumn{1}{c}{15.0573}
			&\multicolumn{1}{c}{2.000}
			&\multicolumn{1}{c}{11.796}
			&\multicolumn{1}{c}{5.7837}
			&\multicolumn{1}{c}{1.959}
			&\multicolumn{1}{c}{201.840}
			\\
			\multicolumn{1}{c}{-0.5}
			&\multicolumn{1}{c}{16.0571}
			&\multicolumn{1}{c}{2.028}
			&\multicolumn{1}{c}{11.693}
			&\multicolumn{1}{c}{6.0217}
			&\multicolumn{1}{c}{1.985}
			&\multicolumn{1}{c}{203.134}
	
	\\		
						
			\hline
			\multicolumn{1}{c}{-1}
			&\multicolumn{1}{c}{6.3643}
			&\multicolumn{1}{c}{1.000}
			&\multicolumn{1}{c}{12.259}
			&\multicolumn{1}{c}{2.5762}
			&\multicolumn{1}{c}{1.502}
			&\multicolumn{1}{c}{535.910}
													\\
			\multicolumn{1}{c}{-1}
			&\multicolumn{1}{c}{7.1212}
			&\multicolumn{1}{c}{1.150}
			&\multicolumn{1}{c}{12.279}
			&\multicolumn{1}{c}{2.9480}
			&\multicolumn{1}{c}{1.548}
			&\multicolumn{1}{c}{438.040}				\\
													
			\multicolumn{1}{c}{-1}
			&\multicolumn{1}{c}{8.7378}
			&\multicolumn{1}{c}{1.400}
			&\multicolumn{1}{c}{12.242}
			&\multicolumn{1}{c}{3.6215}
			&\multicolumn{1}{c}{1.623}
			&\multicolumn{1}{c}{344.180}	
													\\
			\multicolumn{1}{c}{-1}
			&\multicolumn{1}{c}{10.6688}
			&\multicolumn{1}{c}{1.600}
			&\multicolumn{1}{c}{12.109}
			&\multicolumn{1}{c}{4.2775}
			&\multicolumn{1}{c}{1.687}
			&\multicolumn{1}{c}{307.820}	
													\\								
			\multicolumn{1}{c}{-1}
			&\multicolumn{1}{c}{13.8956}
			&\multicolumn{1}{c}{1.788}
			&\multicolumn{1}{c}{11.816}
			&\multicolumn{1}{c}{5.1462}
			&\multicolumn{1}{c}{1.760}
			&\multicolumn{1}{c}{302.630}	
		\end{tabular}
	\end{ruledtabular}
\end{table*}

\begin{figure}
	\centering
	\includegraphics[width=0.5\textwidth]{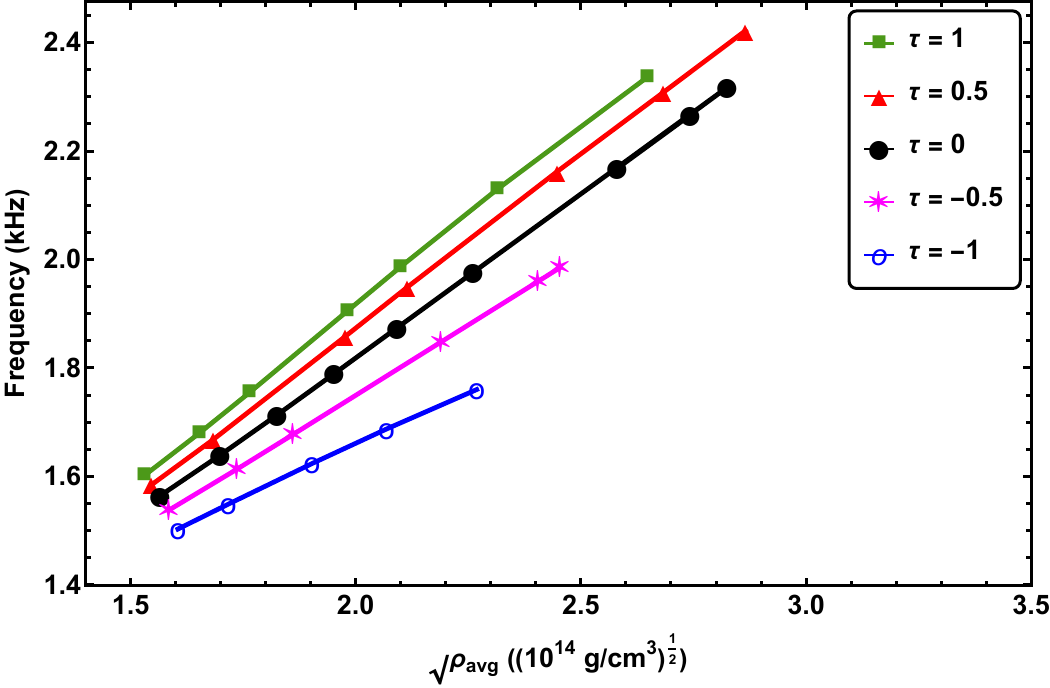}
	\caption{The variation of the f-mode frequency with the square root of the average density of the neutron stars for various values of the anisotropic strength for BSk21 EoS.}
	\label{f-sqrt_rho}
\end{figure}

Next, we investigate the effect of the anisotropy on the properties of the f-mode oscillation of neutron stars. In Fig. \ref{f-sqrt_rho}, we plot the frequency of the f-mode along the abscissa and the square root of the average density along the ordinate. From the calculations within the Newtonian theory, we know that the f-mode frequency is proportional to the square root of the average density of a neutron star \cite{Andersson:2019yve} even in the case with anisotropy \cite{hillebrandt1976anisotropic}. From Fig. \ref{f-sqrt_rho}, we see that this frequency versus the square root of the average density ($\mathcal{F}-\sqrt{\rho_{\rm avg}}$) relation remains unchanged even when full general relativistic calculations are performed. The higher values of the anisotropic strength results in higher values of the slope of the $\mathcal{F}-\sqrt{\rho_{\rm avg}}$ lines. From this figure, it is clear that the f-mode frequency can be written as a functional form as:
	\begin{equation}\label{eq:fmodefreq}
	\mathcal{F}(\rho_{\rm avg},\tau) \approx C \sqrt{\rho_{\rm avg}} \, g(\tau)~,
	\end{equation}
where $C$ is a positive constant of proportionality, and $g(\tau)$ is a monotonically increasing function of $\tau$. 
 
\begin{figure}
	\centering
	\includegraphics[width=0.5\textwidth]{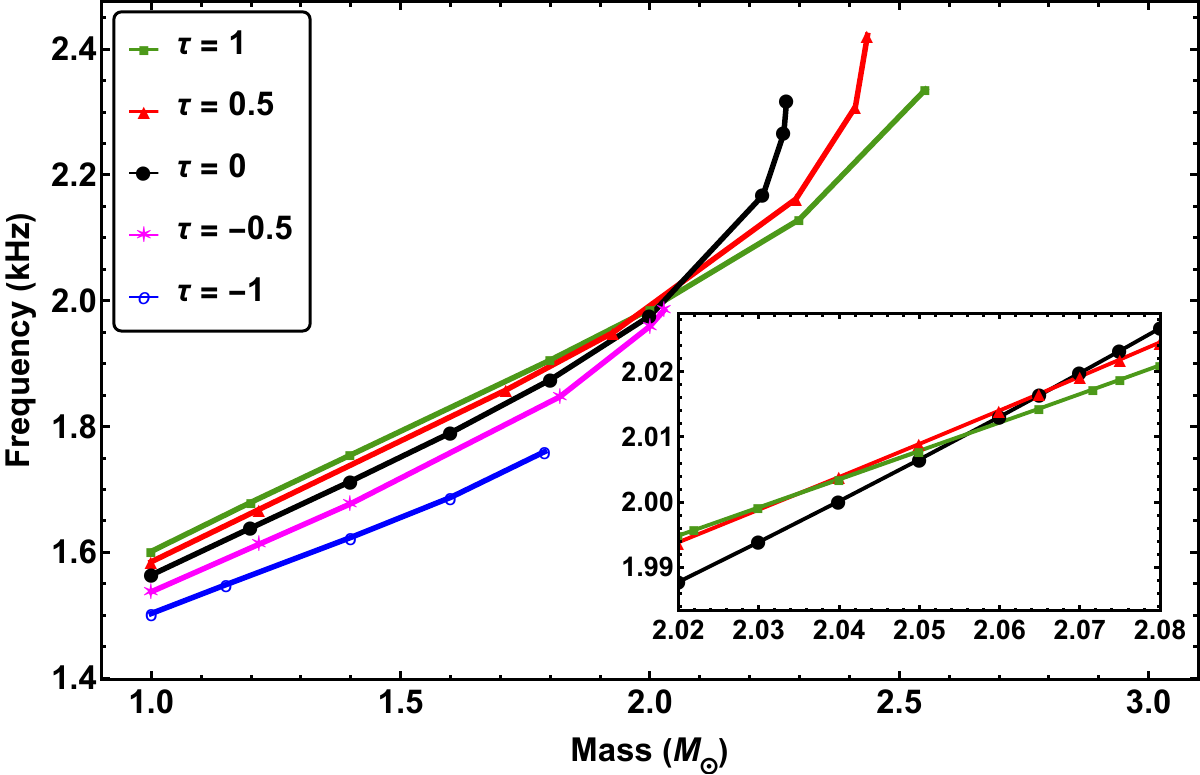}
	\caption{The variation of the f-mode frequency with the mass of neutron stars for various values of the anisotropic strength for BSk21 EoS.}
	\label{f-vs-M}
\end{figure}

In Fig. \ref{f-vs-M}, we plot the frequencies against the mass of the neutron stars for several values of the anisotropic strength. From this figure, we see that, for any given value of the anisotropic strength, neutron stars of higher masses have higher values of the f-mode frequency. For lower masses (less than about $2~{\rm M_\odot}$), the increase of the frequency with the mass is linear, but for massive stars, the frequency increases rapidly with the increase in the value of the mass. However, this non-linear rise of the frequency with the mass is not prominent for negative values of $\tau$. We also see that, for a fixed mass below about $2~{\rm M_\odot}$, a higher value of $\tau$ has a larger frequency, but above that mass, a higher value of $\tau$ corresponds to a smaller value of the frequency. 
The crossing points of two constant-$\tau$ lines vary slightly depending on the values of $\tau$ as shown in the inset, which zooms in on the crossing region. Specifically, the lines for $\tau=1$ and $\tau=0.5$ intersect at $M=2.036 ~{\rm M_{\odot}}$, the lines for  $\tau=1$ and $\tau=0.5$ intersect at $M=2.055 ~{\rm M_{\odot}}$, and the lines for $\tau=0.5$ and $\tau=0$ intersect at $M=2.065 ~{\rm M_{\odot}}$. 

\begin{figure}
	\centering
	\subfloat[\centering The variation of the f-mode frequency with the anisotropic strength for near solar mass neutron stars.]{\includegraphics[width=0.5\textwidth]{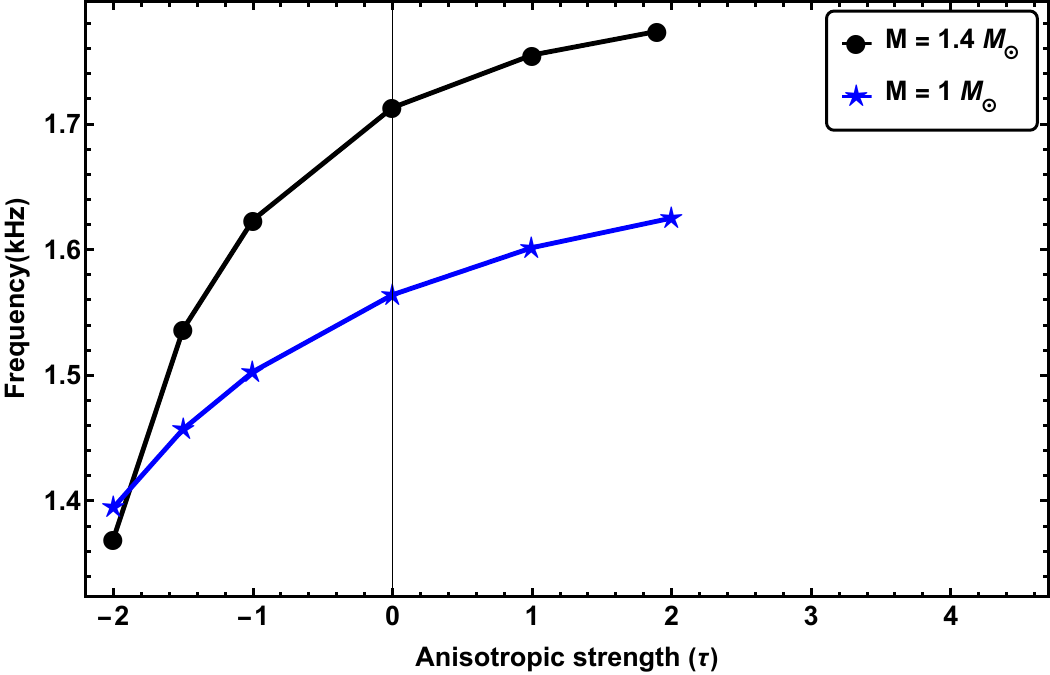}\label{f-vs-tau:subfig1}}\hfill
	\subfloat[\centering The variation of the f-mode frequency with the anisotropic strength for massive neutron stars.]{\includegraphics[width=0.5\textwidth]{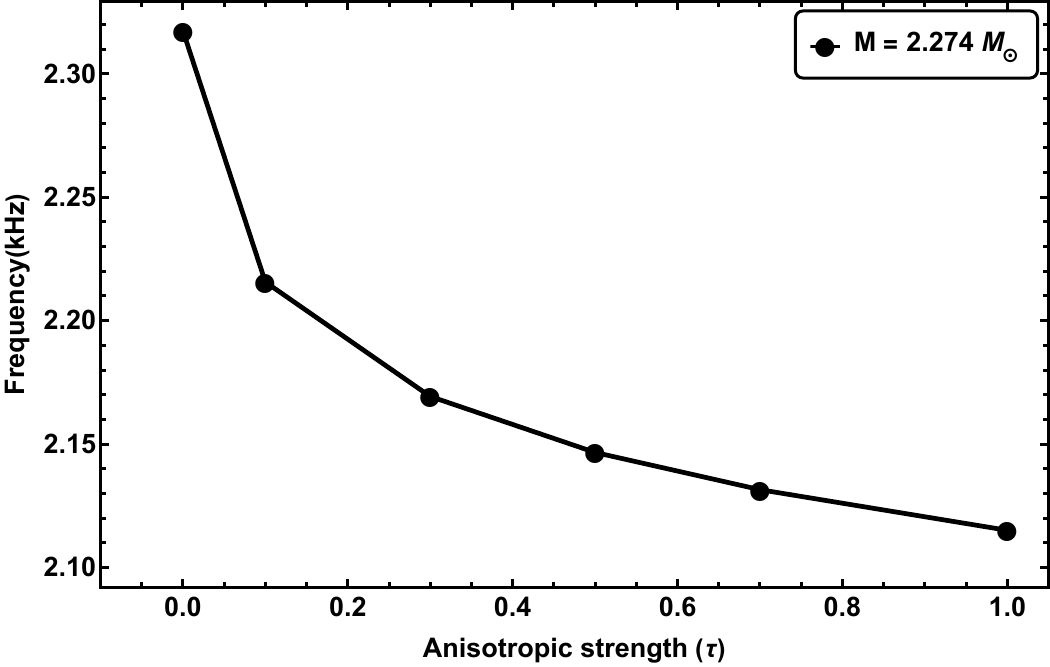}\label{f-vs-tau:subfig2}}.
	\caption{Frequencies of f-modes for various anisotropic strengths for three different masses of neutron stars for BSk21 EoS.} 
	\label{f-vs-tau}
\end{figure}

The rapid increase in the values of the f-mode frequency with the mass for high values of masses (as seen in Fig. \ref{f-vs-M}) can be understood from Eq. (\ref{eq:fmodefreq}) and Figs. \ref{sqrtrhovstau} and \ref{f-sqrt_rho}. Fig. \ref{sqrtrhovstau} demonstrates the fact that for a fixed value of the anisotropic strength, the value of $\sqrt{\rho_{\rm avg}}$ of a neutron star increases with its mass, and this increase becomes faster for higher values of the mass. For example, when $\tau=0$, if the mass of the neutron star increases from $1.0~{\rm M_\odot}$ to $1.4~{\rm M_\odot}$, i.e, $\Delta M = 0.4 ~{\rm M_\odot}$, the value of $\sqrt{\rho_{\rm avg}}$ rises from  $1.56 \times 10^{7} ~{\rm g^{1/2} \, cm^{-3/2}}$ to $1.82 \times 10^{7} ~{\rm g^{1/2} \, cm^{-3/2}}$, i.e., $\Delta \sqrt{\rho_{\rm avg}} = 0.26 \times 10^{7} ~{\rm g^{1/2} \, cm^{-3/2}}$ giving $\Delta \sqrt{\rho_{\rm avg}} / \Delta M   = 0.65 \times 10^{7} ~{\rm g^{1/2} \, cm^{-3/2}  \, M_{\odot}^{-1} } $. On the other hand, keeping $\tau$ fixed at 0, if the mass increases from $1.4~{\rm M_\odot}$ to $2.274~{\rm M_\odot}$, i.e, $\Delta M = 0.874 ~{\rm M_\odot}$, the value of $\sqrt{\rho_{\rm avg}}$ increases from $1.82 \times 10^{7} ~{\rm g^{1/2} \, cm^{-3/2}}$ to $2.82 \times 10^{7} ~{\rm g^{1/2} \, cm^{-3/2}}$, i.e, $\Delta \sqrt{\rho_{\rm avg}} = 1.0  \times 10^{7} ~{\rm g^{1/2} \, cm^{-3/2}}$ giving $ \Delta \sqrt{\rho_{\rm avg}} /\Delta M  = 1.14 \times 10^7 ~{\rm g^{1/2} \, cm^{-3/2} \, M_{\odot}^{-1} }$. In summary, the value of $ \Delta \sqrt{\rho_{\rm avg}} /\Delta M$ increases as the mass increases, a trend true for any fixed value of $\tau$. Furthermore, Fig. \ref{f-sqrt_rho} shows that for a fixed value of $\tau$, the f-mode frequency is proportional to $\sqrt{\rho_{\rm avg}}$. These two facts jointly result in the rapid non-linear rise in the value of the frequency with the increase in the value of the mass  in the high-mass regime (for any constant $\tau$ value), as seen in Fig. \ref{f-vs-M}.

Fig. \ref{f-sqrt_rho} shows that $g(\tau)$ of Eq. (\ref{eq:fmodefreq}) is a monotonically increasing function of $\tau$, and Fig. \ref{sqrtrhovstau} illustrates that in the low-mass regime, the variation of $\sqrt{\rho_{\rm avg}}$ with $\tau$ is minimal. Consequently, in this regime, the monotonically increasing function $g(\tau)$ dominates the behavior of $\mathcal{F}(\rho_{\rm avg}, \tau)$ as defined in Eq. (\ref{eq:fmodefreq}). This explains why, in the low-mass region of Fig. \ref{sqrtrhovstau}, higher values of $\tau$ correspond to higher frequency values for a fixed mass. On the other hand, in the high-mass regime, $\sqrt{\rho_{\rm avg}}$ decreases rapidly with increasing $\tau$ as shown in Fig. \ref{sqrtrhovstau}. Hence, in this regime, $\sqrt{\rho_{\rm avg}}$ has a greater influence on $\mathcal{F}(\rho_{\rm avg}, \tau)$, leading to lower f-mode frequencies for larger values of $\tau$ as observed for higher mass values in Fig. \ref{f-vs-M}.

To understand the impact of anisotropy on the f-mode frequency for a specific mass of a neutron star, in Fig. \ref{f-vs-tau}, we have plotted the frequency with respect to the anisotropic strength for the same three values of the mass that were used in Fig. \ref{sqrtrhovstau}. It is important to note that the mass value $2.274~{\rm M_\odot}$, as used in Fig. \ref{f-vs-tau:subfig2}, is above the threshold mass value, where an increase in $\tau$ leads to a decrease in the mode frequency for a fixed mass, as seen in Fig. \ref{f-vs-M}. In contrast, masses $1.0~{\rm M_\odot}$ and $1.4~{\rm M_\odot}$, as used in Fig. \ref{f-vs-tau:subfig1}, are below this threshold. From Fig. \ref{f-vs-tau:subfig1}, we see the fact that for neutron stars of masses $1~{\rm M_\odot}$ and $1.4~{\rm M_\odot}$, the f-mode frequency increases as the anisotropic strength increases. Comparing the f-mode frequency of an anisotropic star with $\tau = 1.9$ and a mass of $1.4~{\rm M_\odot}$ to that of an isotropic star with the same mass, we observe a $3.62\%$ increase in the frequency. Similarly, comparing the frequency of an anisotropic star with $\tau = -1.9$ and a mass of $1.4~{\rm M_\odot}$ to that of an isotropic star of the same mass, we see a $17.37\%$ decrease in the frequency. The f-mode frequency of a $1.0~\rm{M_\odot}$ star with $\tau = 1.9$ is $3.78 \%$ higher than that of an isotropic star of the same mass. For $\tau = - 1.9$ and $M = 1.0~\rm{M_\odot}$, the frequency decreases by $9.91 \%$ compared to an isotropic star of the same mass. On the other hand, for a neutron star of mass $2.274~\rm{M_\odot}$, the f-mode frequency decreases as the anisotropy inside the star increases. From Fig. \ref{f-vs-tau:subfig2}, we see that the frequency experiences a decrease of approximately $9.56\%$ as the anisotropic strength increases from $0$ to $1$. This characteristic, where an increase in the anisotropic strength leads to a decrease in the f-mode frequency, also exists in the framework of Cowling approximation \cite{Curi:2022nnt}.

\begin{figure}
	\centering
	\includegraphics[width=0.5\textwidth]{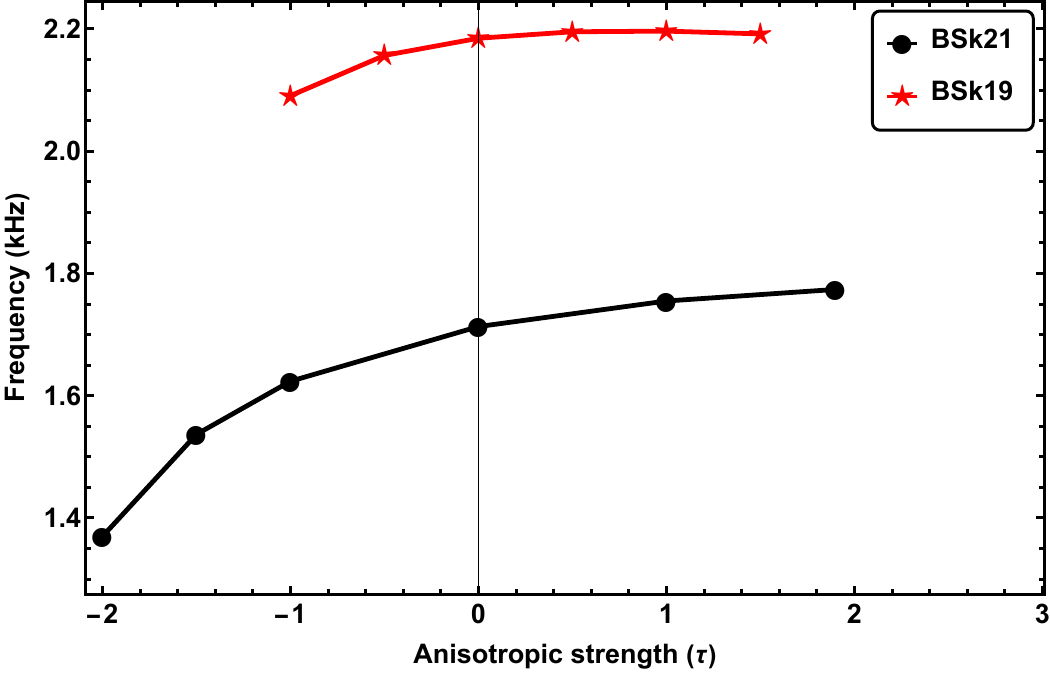}
	\caption{The variation of the f-mode frequency with the anisotropic strength for neutron stars of mass $1.4~\rm M_\odot$ for BSk19 and BSk21 EsoS. }
	\label{f-vs-tau-1921}
\end{figure}

Next, we explore the influence of the stiffness of the EoS on the f-mode frequency. We plot the frequency as a function of the anisotropic strength for both the BSk19 and BSk21 EsoS in Fig. \ref{f-vs-tau-1921}. To facilitate a direct comparison of the stiffness effect, we considered a fixed mass $1.4~{\rm M_\odot}$ for the neutron stars for both of the EsoS. For BSk21 EoS, stable neutron stars of mass $1.4~{\rm M_\odot}$ can exist in the range of the anisotropic strength $-2 \lesssim \tau \lesssim 1.9$, while for BSk19 EoS, this range is  $-1 \lesssim \tau \lesssim 1.5$. The figure indicates that the qualitative behavior of both curves is similar, suggesting a common trend. However, neutron stars with a softer EoS, such as BSk19, exhibit higher oscillation frequencies across the entire range of anisotropy. Additionally, it is noteworthy that the impact of anisotropy is less pronounced for the softer EoS (BSk19) compared to the stiffer EoS (BSk21), particularly when $\tau > 0$.

\begin{figure}
	\centering
	\includegraphics[width=0.5\textwidth]{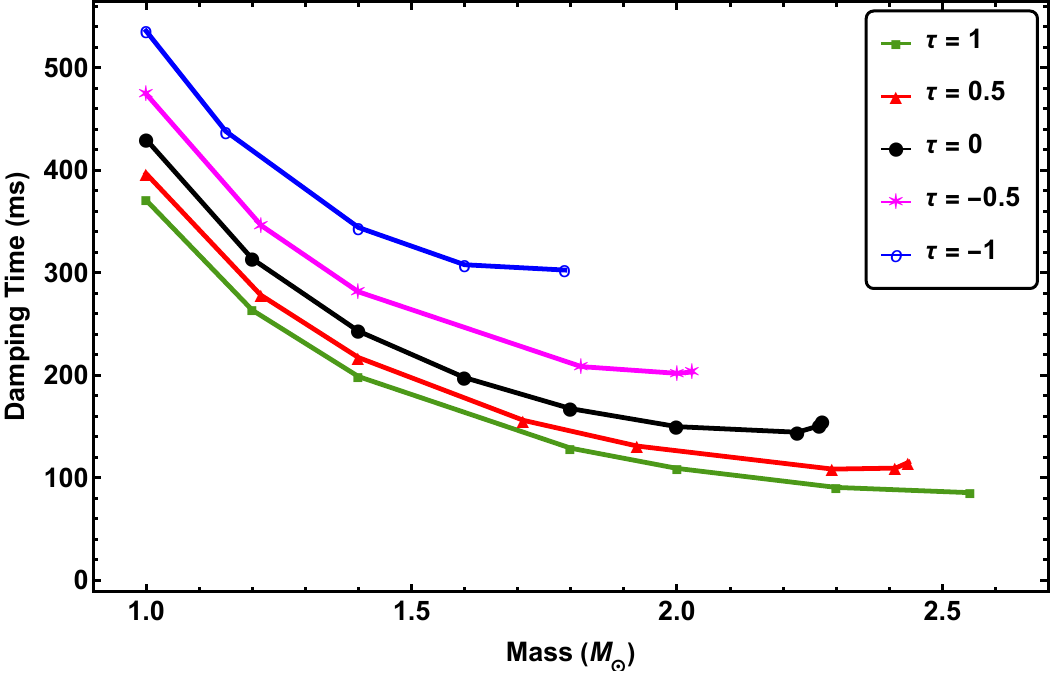}
	\caption{The variation of the damping time with the mass of the neutron stars for various values of the anisotropic strength for BSk21 EoS.}
	\label{T-vs-M}
\end{figure}

\begin{figure}
\centering
\includegraphics[width=0.5\textwidth]{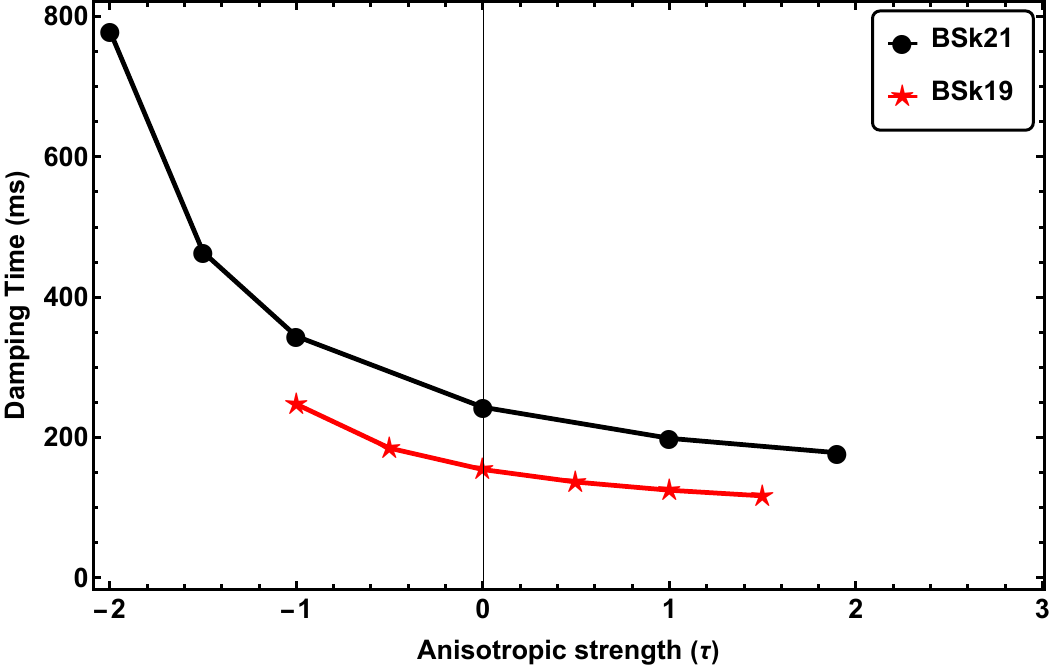}
\caption{The variation of the damping time with the value of the anisotropic strength for neutron stars with the mass of $1.4~{\rm M_\odot}$ for BSk19 and BSk21 EsoS.}
\label{T-vs-tau-1921}
\end{figure}

In addition to the mode frequency, another crucial aspect associated with the mode is the damping time. Hence, next, we investigate the effect of the anisotropy on the damping time of the f-mode oscillation. In Fig. \ref{T-vs-M}, we plot the values of the damping time against the mass of neutron stars for several values of the anisotropic strength. From this figure, we see that the damping time decreases as the mass of the neutron star increases for all values of the anisotropic strength. However, it is worth noting that for isotropic or mildly anisotropic neutron stars, a slight increase in the damping time occurs near the stable maximum mass. 

In Fig. \ref{T-vs-tau-1921}, we plot the values of the anisotropic strength along the abscissa and the values of the damping time along the ordinate for neutron stars of a fixed mass $1.4~{\rm M_\odot}$ for EsoS BSk21 and BSk19. From this figure, we see that the qualitative nature of the two curves is similar, indicating a common trend. However, it is evident that the neutron stars with a softer EoS like the BSk19 EoS, give smaller values of the damping time of oscillations across the entire range of the anisotropy. We also notice the fact that, for a neutron star of mass $1.4~{\rm M_\odot}$ with BSk21 EoS and an anisotropic strength of $\tau = 2$, the damping time experiences a decrease of $28\%$ compared to an isotropic star of the same mass. Conversely, for the same mass and the EoS, as the value of $\tau$ decreases from $0$ to $-2$, we observe a significant increase in the damping time, approximately three times ($300\%$) of that for the isotropic case. For neutron stars of mass $1.4~{\rm M_\odot}$ with BSk19 EoS, as the anisotropic strength increases from $0$ to $1.5$, the damping time decreases by $24\%$. Conversely, as the anisotropic strength decreases from $0$ to $-1$, the damping time increases by $60\%$.

\section{Summary and Conclusion}\label{sec:conclusion}

In this paper, we studied the effects of the anisotropy within neutron stars on the frequency and the damping time of the quasi-normal modes, as compared to those of isotropic neutron stars. To achieve this, we have employed the full framework of general relativity (up to the linear order perturbation), considering both the metric and the fluid perturbations to derive the oscillation equations. Notably, these equations share a similar structure to the isotropic counterparts but contain additional terms that account for the difference between the radial and the tangential pressure and its perturbations. 

In a recent work, \citet{Lau:2024oik}, considered structural anisotropy in addition to pressure anisotropy. This lead to a non-zero perturbation in the strain tensor affecting the perturbation in the density ($\Delta \rho$). As a result, their expressions of $X^{\prime}$ and $W^{\prime}$ have some extra terms. However, both our team and \citet{Lau:2024oik} independently verified that these extra terms have a negligible effect on the numerical values of the frequencies and damping times of f-mode oscillations in non-rotating neutron stars. We plan to investigate the effects of structural anisotropy on non-radial oscillations in rotating neutron stars in future research.

As our focus is on gravitational waves, we consider the specific case of $l = 2$, although the equations are applicable for any value of $l$. Among the various $l=2$ modes, we have focused on studying the fundamental (f) modes of neutron stars, which typically span a frequency range of $1 - 3$ kHz. These modes are relatively easier to excite compared to other modes \cite{Ferrari:2007dd}, and their analysis provides crucial insights into the oscillatory behavior and dynamic characteristics of neutron stars. In our analysis, we have demonstrated that the frequency of the f-mode exhibits a linear relationship with the square root of the average density of neutron stars, for both isotropic and anisotropic neutron stars. Notably, the slope of the linear fit depends on the anisotropic strength $\tau$, it increases with the increase of $\tau$ from zero, and decreases with the decrease of $\tau$ from zero. Such relation was obtained earlier within the Newtonian framework by \citet{hillebrandt1976anisotropic}.

Since the mass of the neutron stars can be determined by observing them in the electromagnetic waves, we have plotted the variations of the f-mode frequency and the damping time with the mass of the star in Fig. \ref{f-vs-M} and Fig. \ref{T-vs-M}, respectively. We have seen that, the qualitative nature of these variations are the same for both the isotropic as well as the anisotropic stars. As shown in Fig. \ref{f-vs-tau}, for neutron stars of masses $1~{\rm M_\odot}$ and $1.4~{\rm M_\odot}$, the frequency increases with the increase in the anisotropic strength. However, for a neutron star of mass $2.274~{\rm M_\odot}$, the frequency decreases as the anisotropic strength decreases. As discussed earlier, this feature is attributed to the rapid decrease of the square root of the average density with the increase of the anisotropic strength for massive neutron stars and very slow decrease of the square root of the average density with the increase of the anisotropic strength for neutron stars of masses near the solar mass, as seen in Fig. \ref{sqrtrhovstau}.

To understand the effect of the stiffness of the EoS on the properties of the f-mode oscillation of anisotropic neutron stars, we have also estimated the frequencies and the damping times using BSk19 EoS for various values of the mass and the anisotropic strength. In Fig. \ref{f-vs-tau-1921}, we showed that for the same value of $\tau$, the value of the f-mode frequency is higher for the softer EoS, but the nature of the change in the frequency due to the change in the anisotropic strength is similar for both EsoS. We have also studied the effect of the anisotropy on the damping time of the f-mode oscillation. From Fig. \ref{T-vs-M}, it is clear that, irrespective of the presence of the anisotropy in the star, the damping time decreases with the increase in the mass of the neutron stars. From Fig. \ref{T-vs-tau-1921}, it is clear that the damping time decreases as $\tau$ increases. This figure also establishes the fact that for the same value of the anisotropic strength, the softer EoS gives smaller values of the damping time than that given by the stiffer EoS.

Our study provides valuable insights into the role of anisotropy in the f-mode oscillations of neutron stars. While early analyses using Newtonian and Cowling approximations offered preliminary views on how anisotropy influences the structure and oscillatory behavior of neutron stars, our current investigation, which accounts for perturbations in both the metric and the fluid, presents a more detailed understanding of f-mode oscillation frequencies.

Moreover, our analysis enables the calculation of the damping times linked to the energy dissipation through gravitational waves. These measurable properties, the f-mode frequency and its damping time, serve as unique signatures of neutron stars, encapsulating essential traits such as the mass, the radius, and the anisotropy. Although our approach employs realistic equations of state, the precise nature of the anisotropy remains approximated due to the limitations in the data availability. To gain deeper insights into the intricate anisotropic pressure, future research must unravel its complexities, enriching our grasp of neutron stars' micro- and macroscopic attributes. Moreover, we have studied the effect of the pressure anisotropy only on the f-mode oscillation of neutron stars. Similar study on the effect of the pressure anisotropy on the properties of other modes of the non-radial oscillations, e.g., the p-modes and the w-modes, of neutron stars will be complementary to this study. Finally, as neutron stars are known as rapid rotators, a more realistic picture about the properties of the anisotropic neutron stars can be obtained by including the effect of rotation in the calculation, which is beyond the scope of the present paper.   

\section{Acknowledgements}
The authors thank the anonymous referee for constructive suggestions and Shu Yan Lau and group for email communications.

\bibliography{main}
\end{document}